\documentclass[10pt]{article}

\usepackage[top=1in, bottom=1.25in, left=1.25in, right=1.25in]{geometry}

\usepackage{setspace}
\usepackage{algorithm}

\usepackage{algpseudocode}

\usepackage{amssymb}

\usepackage{amsthm}

\usepackage{pifont}

\usepackage{amsmath}

\usepackage{adjustbox}

\usepackage{bm}

\usepackage{bbm}

\usepackage{natbib}

\usepackage{graphicx}

\usepackage{booktabs}

\usepackage{blkarray}

\usepackage{titlesec}  

\usepackage{authblk}

\usepackage{xcolor,colortbl}

\titleformat*{\section}{\sffamily\large}
\titleformat*{\subsection}{\sffamily\normalsize}

\newcommand{\mat}[1]{\bm{#1}}
\newcommand{\vect}[1]{\bm{#1}}
\newcommand{\set}[1]{\left\lbrace #1 \right\rbrace}
\newcommand{\T}{\mathsf{T}}

\newcommand{\xmark}{\ding{55}}%

\newcommand\shadecell{\cellcolor{black!10}}

\graphicspath{{Figures/}}

\title{Statistical matching of non-Gaussian data}

\author[1]{Daniel Ahfock}
\author[2,3]{Saumyadipta Pyne}
\author[1]{Geoffrey J. McLachlan}
\affil[1]{School of Mathematics and Physics, University of Queensland, Brisbane, Australia}
\affil[2]{Public Health Dynamics Laboratory, Graduate School of Public Health, University of Pittsburgh,
Pittsburgh, PA, USA}
\affil[3]{Department of Biostatistics, Graduate School of Public Health, University of Pittsburgh, Pittsburgh, PA,
USA}

\date{}                     
\begin{document}
\maketitle

\begin{abstract}
The statistical matching problem is a data integration problem with structured missing data. The general form involves the analysis of multiple datasets that only have a strict subset of variables jointly observed across all datasets. The simplest version involves two datasets, labelled A and B, with three variables of interest $X, Y$ and $Z$. Variables $X$ and $Y$ are observed in dataset A and variables $X$ and $Z$ are observed in dataset $B$. Statistical inference is complicated by the absence of joint $(Y, Z)$ observations. Parametric modelling can be challenging due to identifiability issues and the difficulty of parameter estimation. We develop computationally feasible procedures for the statistical matching of non-Gaussian data using suitable data augmentation schemes and identifiability constraints. Nearest-neighbour imputation is a common alternative technique due to its ease of use and generality. Nearest-neighbour matching is based on a conditional independence assumption that may be inappropriate for non-Gaussian data. The violation of the conditional independence assumption can lead to improper imputations. We compare model based approaches to nearest-neighbour imputation on a number of flow cytometry datasets and find that the model based approach can address some of the weaknesses of the nonparametric nearest-neighbour technique.  
\end{abstract}

%



%

\section{Introduction}
The statistical matching problem is a data integration scenario where the set of available datasets shares only a strict subset of variables. The pattern of missingness is such that some dimensions are wholly unobserved in each dataset \citep{dorazio_2006_statistical}. The simplest case involves two datasets, dataset A and dataset B, and three groups of variables, labelled $\vect{X}$, $\vect{Y}$ and $\vect{Z}$. We assume the data generating process is governed by a parametric model $f(\vect{x}, \vect{y}, \vect{z}; \vect{\theta})$, for some parameter $\vect{\theta} \in \Omega \subset \mathbb{R}^p$. Dataset A contains $n_{A}$ independent observations from the marginal distribution $f(\vect{x}, \vect{y}; \vect{\theta})$ and dataset $B$ consists of $n_{B}$ independent observations from the marginal distribution $f(\vect{x}, \vect{z}; \vect{\theta})$. Table \ref{tab:simple} represents the missing data pattern. A common objective in the statistical matching problem is to impute the missing observations in each dataset so complete data techniques can be used in downstream analyses \citep{rassler_2002_statistical}. The statistical matching problem occurs in flow cytometry analysis due to technological limitations, and there has been recent interest in developing appropriate statistical methods for  integrating cytometry datasets \citep{pedreira_2008_generation, oneill_2015_deep}. Most previous work on the statistical matching problem for continuous data assumes a multivariate normal model. Flow cytometry data typically have characteristics that make a multivariate Gaussian model inappropriate. Observations are from a number of heterogeneous cell subpopulations, and observations within a cell subtype often exhibit skewness and heavy tails. Finite mixtures of skew-normal and skew-$t$ distributions have shown excellent performance in the analysis of flow cytometry data \citep{pyne2009automated, lee_2016_finite}. We study theoretical aspects of the statistical matching of non-Gaussian data and propose new computationally efficient algorithms for the statistical matching of skew-normal data, mixtures of Gaussians and mixtures of skew-normal distributions.

\begin{table}
\centering
\begin{tabular}{lrrr}
\toprule
 Dimensions & $\vect{X}$ & $\vect{Y}$ & $\vect{Z}$ \\
\midrule
Dataset A & \checkmark & \checkmark & \xmark \\
Dataset B & \checkmark & \xmark & \checkmark \\
\bottomrule
\end{tabular}
\caption{Missing data pattern in the statistical matching problem.}
\label{tab:simple}
\end{table}

A central issue in the statistical matching problem is that the lack of joint observations on the $\vect{Y}$ and $\vect{Z}$ variables renders most statistical models nonidentifiable. For example, if $f(\vect{x}, \vect{y}, \vect{z} ; \vect{\theta})$ is a multivariate normal distribution, the $\vect{Y}, \vect{Z}$ covariance parameters are not identifiable. In a parametric framework, it is common to adopt restrictions on the parameters so that the generative model remains identifiable. Under identification constraints it is possible to impute the missing data using a parametric modelling strategy. Let $\widehat{\vect{\theta}}$ denote the maximum likelihood estimate of $\vect{\theta}$. The missing  $\vect{Z}$ values in dataset A can be sampled from  $f(\vect{z}| \vect{x},\vect{y}; \widehat{\vect{\theta}})$, and the missing $\vect{Y}$ values in dataset B can be sampled from $f(\vect{y}| \vect{x},\vect{z}; \widehat{\vect{\theta}})$.

Parametric strategies for Gaussian data typically involve reparameterising the model so that the likelihood can then be factored into a mathematically tractable form \citep{dorazio_2006_statistical}. Reparameterisation and factorisation are powerful strategies in missing data problems, as it can greatly simplify maximum likelihood estimation \citep[][Chapter 7]{little_2002_statistical}. The reparameterisation for the statistical matching of Gaussian data involves expressing the observed-data likelihood in terms of a multiple response regression model. We show how the regression model specification can be extended to cover skew-normal distributions, mixtures of Gaussians, and mixtures of skew-normal distributions through appropriate data augmentation schemes. Factorisation of the complete-data log likelihood allows the EM algorithm \citep{dempster_maximum_1977} to be used for computationally efficient parameter estimation. 

An alternative method for imputation is nearest-neighbour matching, a nonparametric method based on the assumption that the $\vect{Y}$ and $\vect{Z}$ variables are conditionally independent given the $\vect{X}$ variables. Nearest-neighbour imputation has been widely applied in the statistical matching problem due to its ease of implementation and generality \citep{dorazio_2006_statistical}. In particular, nearest-neighbour matching has been advocated for the file matching of flow cytometry data \citep{pedreira_2008_generation}. An important finding is that the $\vect{Y}$ and $\vect{Z}$ variables are likely to be conditionally dependent given the $\vect{X}$ variables if the generative model is a mixture model or a skewed distribution. The violation of the conditional independence assumption can lead to nearest-neighbour imputation exhibiting undesirable behaviour. 

The conditional independence assumption can be adopted for any data generating process, and is the dominant approach to resolve the pathological missing data pattern in the statistical matching problem. This conditional independence assumption is difficult to justify for a range of latent variable models, and model based approaches can impose more coherent constraints. We compare nearest-neighbour matching to model based imputation on a number of real datasets. The model based approach showed a large improvement over the nearest-neighbour method, however there can be issues when the identification constraints are violated.  Parametric imputation strategies are a flexible and computationally feasible option for statistical matching.

\section{Background}
\subsection{Notation}
We assume $n_{A}$ observations in dataset A and $n_{B}$ observations in dataset B, for a total of $n=n_{A}+ n_{B}$ records. These datasets are combined to form a single data matrix, where rows 1 to $n_{A}$  contain the records from dataset $A$ on the $\vect{X}$ and $\vect{Y}$ variables and rows  $n_{B}$ rows contain the observations from dataset $B$  on the $\vect{X}$ and $\vect{Z}$ variables. Let $d_{X}$, $d_{Y}$ and $d_{Z}$ give the dimension of the $\vect{X}, \vect{Y}$ and $\vect{Z}$ variables respectively and set $d=d_{X}+d_{Y}+d_{Z}$. 

We assume the data generating process consists of $n$ independently and identically distributed observations from the parametric model
 $(\vect{X}_{i}^{\T}, \vect{Y}_{i}^{\T}, \vect{Z}_{i}^{\T})^{\T} \sim f(\vect{x}_{i}, \vect{y}_{i}, \vect{z}_{i}; \vect{\theta})$ for $i=1, \ldots, n$. Let $\vect{x}_{i}$, $\vect{y}_{i}$ and $\vect{z}_{i}$ represent the realised values for the $\vect{X}_{i}$, $\vect{Y}_{i}$ and $\vect{Z}_{i}$ random vectors respectively for $i=1, \ldots, n$. Observations $\vect{z}_{i}$ are missing for $i=1, \ldots, n_{A}$. Observations $\vect{y}_{i}$ are missing for $j=n_{A}+1, n_{A}+2, \ldots, n_{A}+n_{B}$. The joint $n \times d$ data matrix of interest is represented in the display \eqref{eq:dataset_structure}. Observed values are shaded, with white cells indicating missing values in the joint data matrix. Let $\vect{x}_{\text{obs}}=(\vect{x}_{1}, \vect{x}_{2}, \ldots, \vect{x}_{n})$ represent the observed $\vect{X}$ values over the $n$ records. Let $\vect{x}_{\text{obs}}^{A}=(\vect{x}_{1}, \ldots, \vect{x}_{n_{A}})$ represent the $n_{A}$ observed $\vect{X}$ values on dataset $A$.

\begin{equation}\label{eq:dataset_structure}
  \left(\begin{array}{cccc}
\shadecell \vect{x}_{1} & \shadecell \vect{y}_{1} & \vect{z}_{1} \\
 \shadecell   \vdots & \shadecell \vdots & \vdots \\
 \shadecell    \vect{x}_{n_{A}} & \shadecell \vect{y}_{n_{A}} & \vect{z}_{n_{A}}\\
  \shadecell   \vect{x}_{n_{A}+1} & \vect{y}_{n_{A}+1} & \shadecell \vect{z}_{n_{A}+1}\\
 \shadecell    \vdots & \vdots & \shadecell \vdots \\
   \shadecell  \vect{x}_{n_{A}+n_{B}} & \vect{y}_{n_{A}+n_{B}} &\shadecell  \vect{z}_{n_{A}+n_{B}}
  \end{array}\right)
\end{equation}
 Let $\vect{x}_{\text{obs}}^{B}=(\vect{x}_{n_{A}+1}, \ldots, \vect{x}_{n_{A}+n_{B}})$ denote the $n_{B}$ observed $\vect{X}$ values in dataset ${B}$. Let $\vect{y}_{\text{obs}}=(\vect{y}_{1}, \ldots, \vect{y}_{n_{A}})$ represent the $n_{A}$ observed values $\vect{Y}$ variables in dataset $A$, and let $\vect{z}_{\text{obs}}=(\vect{z}_{n_{A}+1}, \ldots, \vect{z}_{n_{A}+n_{B}})$ denote the $n_{B}$ observed $\vect{Z}$ values in dataset $B$. Likewise, let $\vect{z}_{\text{mis}}=(\vect{z}_{1}, \ldots, \vect{z}_{n_{A}})$  represent the $n_{A}$ missing $\vect{Z}$ values in dataset $A$, and let $\vect{y}_{\text{mis}}=(\vect{y}_{n_{A}+1}, \ldots, \vect{y}_{n_{A}+n_{B}})$ represent the $n_{B}$ missing  $\vect{Y}$ values in dataset $B$. 

\subsection{Gaussian data}
\label{subsec:gaussian_matching}
The statistical matching problem has been explored in depth under the assumption of a multivariate Gaussian generative model. Suppose that for $i=1, \ldots, n$ we have observations from the multivariate normal distribution
\begin{align}
\begin{bmatrix} \bm{X}_{i} \\ \bm{Y}_{i} \\ \bm{Z}_{i} \end{bmatrix} &\sim N\left(
\vect{\mu} = \begin{bmatrix} \bm{\mu}_X \\ \bm{\mu}_Y \\ \bm{\mu}_Z \end{bmatrix}, {\Sigma} = 
\begin{bmatrix}
{\Sigma}_{XX} & {\Sigma}_{XY} & {\Sigma}_{XZ} \\
{\Sigma}_{YX} & {\Sigma}_{YY} & {\Sigma}_{YZ} \\
{\Sigma}_{ZX} &  {\Sigma}_{ZY} &{\Sigma}_{ZZ}
\end{bmatrix} \right), \label{eq:gaussian_partition}
\end{align}
where the mean and covariance parameters have been partitioned in an obvious fashion. With the missing data pattern represented in \eqref{eq:dataset_structure}, the only non-identifiable parameter is $\Sigma_{YZ}$. The most common identifiability constraint is that $\Sigma_{YZ} = {\Sigma}_{YX}{\Sigma}_{XX}^{-1}{\Sigma}_{XZ}$. 

Parameter estimation under the identification constraint $\Sigma_{YZ} ={\Sigma}_{YX}{\Sigma}_{XX}^{-1}{\Sigma}_{XZ}$ has a long history in the  literature \citep{lord_1955_estimation, anderson_1957_maximum, moriarity_2003_note}. As discussed in  \citet[Chapter 2]{dorazio_2006_statistical}, the multivariate normal statistical matching problem has a useful connection to linear regression modelling that can be used to obtain closed form maximum likelihood estimates. The conditional distribution of $\vect{Y}_{i}$ given $\vect{X}_{i}=\vect{x}_{i}$ can be represented as a regression model
\begin{align}
\vect{Y}_{i} | \vect{X}_{i}=\vect{x}_{i} & \sim N(\vect{\alpha}_{Y} +\vect{\beta}_{Y}\vect{x}_{i}, \Omega_{Y}),\label{eq:y_regression_mvn}
\end{align}
for $i=1, \ldots, n$ where
\begin{align}
\vect{\beta}_{Y} &=  \Sigma_{YX}\Sigma_{XX}^{-1}, \label{eq:beta_Y_definition} \\
\vect{\alpha}_{Y} &= \vect{\mu}_{Y} - \vect{\beta}_{Y} \vect{\mu}_{X}, \label{eq:alpha_Y_definition}  \\
\Omega_{Y} &=  \Sigma_{YY} - \Sigma_{YX}\Sigma_{XX}^{-1}\Sigma_{XY}. \label{eq:omega_Y_definition}
\end{align}
The same holds for the conditional distribution of $\vect{Z}_{i}$ given $\mat{X}_{i}=\vect{x}_{i}$
\begin{align}
\vect{Z}_{i} | \vect{X}=\vect{x}_{i} & \sim N(\vect{\alpha}_{Z} +\vect{\beta}_{Z}\vect{x}_{i}, \Omega_{Z}),  \label{eq:z_regression_mvn}
\end{align}
for $i=1, \ldots, n$ where
\begin{align}
\vect{\beta}_{Z} &=  \Sigma_{ZX}\Sigma_{XX}^{-1}, \label{eq:beta_Z_definition} \\
\vect{\alpha}_{Z} &= \vect{\mu}_{Z} - \vect{\mu}_{X}^{\mathsf{T}}\vect{\beta}_{Z}, \label{eq:alpha_Z_definition} \\
\Omega_{Z} &=  \Sigma_{ZZ} - \Sigma_{ZX}\Sigma_{XX}^{-1}\Sigma_{XZ}. \label{eq:omega_Z_definition} 
\end{align}
Now let $\vect{\eta} = (\vect{\eta}_{X}, \vect{\eta}_{Y}, \vect{\eta}_{Z})$, where
$\vect{\eta}_{X} = (\vect{\mu}_{X}, \Sigma_{XX}), 
\vect{\eta}_{Y} = (\vect{\alpha}_{Y}, \vect{\beta}_{Y}, \Omega_{Y})$ and 
$\vect{\eta}_{Z} = (\vect{\alpha}_{Z}, \vect{\beta}_{Z}, \Omega_{Z})$. The benefit of the regression parameterisation is that that the likelihood for each observation factors into three components with distinct parameter blocks. We can write $f(\vect{x}_{i}, \vect{y}_{i}, \vect{z}_{i} ; \vect{\eta}) = f(\vect{x}_{i}; \vect{\eta}_{X})f(\vect{y}_{i}|\vect{x}_{i}; \vect{\eta}_{Y})f(\vect{z}_{i}| \vect{x}_{i}; \vect{\eta}_{Z})$. The observed-data likelihood can be expressed as a product of three terms,
\begin{align}
f(\vect{x}_{\text{obs}}, \vect{y}_{\text{obs}}, \vect{z}_{\text{obs}} ; \vect{\eta}) &= \prod_{i=1}^{n_{A}}f(\vect{x}_{i}, \vect{y}_{i}; \vect{\eta})\prod_{k=n_{A}+1}^{n}f(\vect{x}_{k}, \vect{z}_{k} ; \vect{\eta}) \\
&= \prod_{i=1}^{n}f(\vect{x}_{i};\vect{\eta}_{X})\prod_{j=1}^{n_{A}}f(\vect{y}_{j}|\vect{x}_{j}; \vect{\eta}_{Y})\prod_{k=n_{A}+1}^{n}f(\vect{z}_{k} |\vect{x}_{k} ; \vect{\eta}_{Z}) \\
&= f(\vect{x}_{\text{obs}}; \vect{\eta}_{X})f(\vect{y}_{\text{obs}} | \vect{x}_{\text{obs}}^{A}; \vect{\eta}_{Y})f(\vect{z}_{\text{obs}}|\vect{x}_{\text{obs}}^{B}; \vect{\eta}_{Z}). \label{eq:likelihood_factorisation}
\end{align}
The first likelihood block $f(\vect{x}_{\text{obs}}; \vect{\eta}_{X})$ involves the $n$ observed $X$ variables. The second likelihood block $f(\vect{y}_{\text{obs}} | \vect{x}_{\text{obs}}^{A}; \vect{\eta}_{Y})$ involves the conditional likelihood for dataset A and the third likelihood block  $f(\vect{z}_{\text{obs}}|\vect{x}_{\text{obs}}^{B}; \vect{\eta}_{Z})$ involves the conditional likelihood for dataset B. Maximisation of the observed-data likelihood
\begin{align}
f(\vect{x}_{\text{obs}}, \vect{y}_{\text{obs}}, \vect{z}_{\text{obs}} ; \vect{\eta}) = f(\vect{x}_{\text{obs}}; \vect{\eta}_{X})f(\vect{y}_{\text{obs}} | \vect{x}_{\text{obs}}^{A}; \vect{\eta}_{Y})f(\vect{z}_{\text{obs}}|\vect{x}_{\text{obs}}^{B}; \vect{\eta}_{Z}) \label{eq:normal_observed_likelihood}
\end{align}
is straightforward, as we now have three separate maximisation problems over $\vect{\eta}_{X}$, $\vect{\eta}_{Y}$ and $\vect{\eta}_{Z}$. Both datasets are used to estimate  $\vect{\eta}_{X}$ using the fully observed $\vect{X}$ variables. Dataset A is used to estimate $\vect{\eta}_{Y}$  and dataset B is used to estimate $\vect{\eta}_{Z}$.

Using the regression specifications \eqref{eq:y_regression_mvn} and \eqref{eq:z_regression_mvn}, we can express  the likelihoods $f(\vect{y}_{\text{obs}} | \vect{x}_{\text{obs}}^{A}; \vect{\eta}_{Y})$ and $f(\vect{z}_{\text{obs}}|\vect{x}_{\text{obs}}^{B}; \vect{\eta}_{Z})$ as multiple response regression models.  Recall the data structure in  display \eqref{eq:dataset_structure}. Let $\vect{Y}_{A}$ give the $n_{A} \times d_{Y}$ matrix of responses from dataset A. Row $i$ in $\mat{Y}_{A}$ is given by $\vect{y}_{i}^{\mathsf{T}}$ for $i=1, \ldots, n_{A}$. Let $\vect{Z}_{B}$ give the  $n_{B} \times d_{Z}$ from dataset B. Row $i$ in $\mat{Z}_{B}$ is given by $\vect{z}_{i}^{\mathsf{T}}$ for $i=n_{A}+1, \ldots, n_{A}+n_{B}$. Let $\mat{B}_{A}$ represent the design matrix for dataset A and let $\mat{B}_{B}$ give the design matrix for dataset $B$. Specifically,
\begin{align}
\mat{B}_{A} &= \begin{bmatrix}
1 & \vect{x}_{1}^{\mathsf{T}} \\
1 &\vect{x}_{2}^{\mathsf{T}} \\
\vdots \\
1 &\vect{x}_{n_{A}}^{\mathsf{T}}
\end{bmatrix}, 
\mat{B}_{B} = \begin{bmatrix}
1 &\vect{x}_{n_{A}+1}^{\mathsf{T}} \\
1 &\vect{x}_{n_{A}+2}^{\mathsf{T}} \\
\vdots \\
1 &\vect{x}_{n_{A}+n_{B}}^{\mathsf{T}}
\end{bmatrix}. \label{eq:design_matrix_normal_app}
\end{align}
Let $\Gamma_{A}$ and $\Gamma_{B}$ contain the regression coefficients for \eqref{eq:y_regression_mvn} and \eqref{eq:z_regression_mvn} respectively: $\Gamma_{A}=[\vect{\alpha}_{Y} \ \vect{\beta}_{Y}]^{\T}$ and $\Gamma_{B}=[\vect{\alpha}_{Z} \ \vect{\beta}_{Z}]^{\T}$. We will use the matrix normal distribution introduced by Dawid (1981) to specify the regression models compactly. A random $n \times p$ matrix $\mat{M}$ is said to have the matrix normal distribution $MN(\mat{I}_{n}, \Sigma)$ if each row is a draw from a  $p$-variate normal distribution $N(\vect{0}, \Sigma)$. The conditional regression models for each dataset can be written as
\begin{align}
\mat{Y}_{A} &= \mat{B}_{A}\mat{\Gamma}_{A} + \vect{\epsilon}_{A},  \label{eq:joint_y_regression_model}\\
\mat{Z}_{B} &= \mat{B}_{B}\mat{\Gamma}_{B} + \vect{\epsilon}_{B}, \label{eq:joint_z_regression_model}
\end{align}
where $\vect{\epsilon}_{A} \sim MN(\vect{I}_{n_{A}}, \Omega_{Y})$ and $\vect{\epsilon}_{B} \sim MN(\vect{I}_{n_{B}}, \Omega_{Z})$. Equation 
\eqref{eq:joint_y_regression_model} is a representation of the likelihood factor $f(\vect{y}_{\text{obs}} | \vect{x}_{\text{obs}}^{A}; \vect{\eta}_{Y})$ and equation \eqref{eq:joint_z_regression_model} is a representation of the likelihood factor $f(\vect{z}_{\text{obs}}|\vect{x}_{\text{obs}^{B}}; \vect{\eta}_{Z})$ . The sufficient statistics for the regressions are  $\mat{Y}_{A}^{\T}\mat{Y}_{A}, \mat{Z}_{B}^{\T}\mat{Z}_{B}, \mat{B}_{A}^{\T}\mat{Y}_{A}, \mat{B}_{B}^{\T}\mat{Z}_{B}, \mat{B}_{A}^{\T}\mat{B}_{A}$ and $\mat{B}_{B}^{\T}\mat{B}_{B}$. The maximum likelihood estimators of the regression parameters $\Gamma_{A}=[\vect{\alpha}_{Y} \ \vect{\beta}_{Y}]^{\T}$ and $\Gamma_{B}=[\vect{\alpha}_{Z} \ \vect{\beta}_{Z}]^{\T}$ are given by
\begin{align*}
    \widehat{\Gamma}_{A} &= (\mat{B}_{A}^{\T}\mat{B}_{A})^{-1}\mat{B}_{A}^{\T}\mat{Y}_{A}, \\
    \widehat{\Gamma}_{B} &= (\mat{B}_{B}^{\T}\mat{B}_{B})^{-1}\mat{B}_{B}^{\T}\mat{Z}_{B}.    
\end{align*}
The maximum likelihood estimators of the error covariance matrices $\Omega_{Y}$ and $\Omega_{Z}$ are given by
\begin{align*}
\widehat{\Omega}_{Y} &=\dfrac{1}{n_{A}} (\mat{Y}_{A}-\mat{B}_{A}\widehat{\Gamma}_{A})^{\T}(\mat{Y}_{A}-\mat{B}_{A}\widehat{\Gamma}_{A}), \\
\widehat{\Omega}_{Z} &= \dfrac{1}{n_{B}}(\mat{Z}_{B}-\mat{B}_{B}\widehat{\Gamma}_{B})^{\T}(\mat{Z}_{B}-\mat{B}_{B}\widehat{\Gamma}_{B}).
\end{align*}
These results follow from general results on multiple response regression models, for example see \citet[][Chapter 10]{rencher_methods_2012}. The maximum likelihood estimators of the $\vect{\eta}_{X}$ parameters are given by the sample mean and the sample covariance of the $n$ observed $\vect{X}$ values $\widehat{\vect{\mu}}_{X} = \dfrac{1}{n}\sum_{i=1}^{n}\vect{x}_{i}$, $
\widehat{\Sigma}_{XX} = \dfrac{1}{n} \sum_{i=1}^{n}(\vect{x}_{i}-\widehat{\vect{\mu}}_{X})(\vect{x}_{i}-\widehat{\vect{\mu}}_{X})^{\T}$. Inverting the transformations defined in equations \eqref{eq:alpha_Y_definition} to \eqref{eq:omega_Z_definition} gives the  maximum likelihood estimates of the original parameters $\vect{\mu}$ and $\vect{\Sigma}$. We thus obtain closed form maximum likelihood estimators under the identification restriction $\Sigma_{YZ} ={\Sigma}_{YX}{\Sigma}_{XX}^{-1}{\Sigma}_{XZ}$. 

\section{Extensions} 
\subsection{Overview}
A parametric approach in the statistical matching problem requires the identification of suitable identifiability constraints, and a feasible procedure for maximum likelihood estimation. One  approach is to introduce the missing data $\vect{y}_{\text{obs}}$, $\vect{z}_{\text{obs}}$ into the model as latent variables. In general, we can form the complete-data likelihood $f(\vect{x}_{\text{obs}},\vect{y}_{\text{obs}}, \vect{z}_{\text{obs}}, \vect{y}_{\text{mis}}, \vect{z}_{\text{mis}}; \vect{\theta})$ and define appropriate EM iterations for parameter estimation. \cite{lee2011statistical} consider the statistical matching of flow cytometry data with mixtures of PCA models and take this approach. Introduction of the missing observations $\vect{y}_{\text{mis}}$ and $\vect{z}_{\text{mis}}$ into the complete data log-likelihood can be computationally demanding on flow cytometry datasets with a large number of observations. Due to the non-identifiability of the model, the fitted mixture model is only used to assign cluster labels to the observations. \citeauthor{lee2011statistical} propose to use nearest-neighbour matching  within each group of labelled points. 

We propose a different method, where we introduce appropriate parameter constraints so that $\vect{y}_{\text{mis}}$ and $\vect{z}_{\text{mis}}$ do not need to be included in the complete-data log likelihood. This is to obtain a more computationally efficient EM algorithm, and to avoid the identifiability issues with fitting a completely unconstrained model. Secondly, we propose to impute the missing data using the constrained fitted model. The strategy involves extending the linear regression connection that was used for Gaussian data. 
\subsection{Skew-normal}
\label{subsec:sn_parametric}
There are many forms of the skew-normal distribution, we will work with the same multivariate version as in \citet{pyne2009automated}. For compact notation, let the vector $\vect{w}_{i}$ represent the vector of joint observations $\vect{w}_{i}=(\vect{x}_{i}^{\mathsf{T}}, \vect{y}_{i}^{\mathsf{T}}, \vect{z}_{i}^{\mathsf{T}})^{\mathsf{T}}$. The density of the skew-normal distribution is
\begin{align}
f(\vect{x}_{i}, \vect{y}_{i}, \vect{z}_{i}; \vect{\mu}, \Sigma, \vect{\delta}) &= 2\phi_{p}(\vect{x}_{i}, \vect{y}_{i}, \vect{z}_{i}; \vect{\mu}, \Lambda)\Phi(\alpha^{\T}(\vect{w}_{i}-\vect{\mu})), \label{eq:skew_normal_density}
\end{align}
where $\Lambda = \Sigma + \vect{\delta}\vect{\delta}^{\mathsf{T}}$ and  $\alpha^{\T} = \vect{\delta}^{\mathsf{T}}\Lambda^{-1}/(1-\vect{\delta}^{\mathsf{T}}\Lambda^{-1}\vect{\delta})$.  Let $TN(\mu, \sigma^2, a)$ denote a lower truncated normal distribution where $\mu$ and $\sigma^2$ give the mean and variance of the underling normal distribution and $a$ gives the lower truncation bound. We say $U \sim TN(\mu, \sigma^2, a)$ if $U\overset{d}{=}[W|W >a]$ where $W \sim N(\mu, \sigma^2)$. The model \eqref{eq:skew_normal_density} has a hierarchical representation,
\begin{align}
\begin{bmatrix}
\vect{X}_{i} \\
\vect{Y}_{i} \\
\vect{Z}_{i}
\end{bmatrix} &= \begin{bmatrix}
\vect{\mu}_{X} \\
\vect{\mu}_{Y} \\
\vect{\mu}_{Z}
\end{bmatrix} + \begin{bmatrix}
\vect{\delta}_{X} \\
\vect{\delta}_{Y} \\
\vect{\delta}_{Z}
\end{bmatrix}{U}_{i}  + \vect{V}, \label{eq:skew_normal_latent_model} 
\end{align}
where ${U}_{i} \sim TN(0, 1, 0)$ and $\vect{V}_{i}\sim N(\vect{0}, \Sigma)$, where $\vect{\mu}$ and $\Sigma$ are partitioned as in \eqref{eq:gaussian_partition}.
The only non-identifiable parameter is $\Sigma_{YZ}$. 
The regression model specification also enables efficient inference for the skew-normal distribution under the identification restriction $\Sigma_{YZ} ={\Sigma}_{YX}{\Sigma}_{XX}^{-1}{\Sigma}_{XZ}$. The statistical matching problem for skew-normal data can be expressed in terms of a multiple response regression model with latent unobserved variables in the conditional mean function. Using the hierarchical model \eqref{eq:skew_normal_latent_model}, conditional on knowing $U_{i}=u_{i}$, it holds that,
 \begin{align}
\left. \begin{bmatrix}
\vect{X}_{i} \\
\vect{Y}_{i} \\
\vect{Z}_{i}
\end{bmatrix} \right\vert U_{i}=u_{i} &\sim \begin{bmatrix}
\vect{\mu}_{X} +\vect{\delta}_{X}u_{i} \\
\vect{\mu}_{Y}+\vect{\delta}_{Y}u_{i} \\
\vect{\mu}_{Z}+\vect{\delta}_{Z}u_{i}
\end{bmatrix}  + \vect{V}_{i}, \label{eq:skew_normal_latent_conditional} 
\end{align}
where $\vect{V}_{i}=N(\vect{0}, \Sigma)$, for $i=1, \ldots, n$. Let $\vect{\beta}_{Y} =  \Sigma_{YX}\Sigma_{XX}^{-1}$ and $\vect{\beta}_{Z} =  \Sigma_{ZX}\Sigma_{XX}^{-1}$. The conditional mean of $\vect{Y}_{i}$ given $\vect{X}_{i}$ and ${U}_{i}$, and the conditional mean of $\vect{Z}_{i}$ given $\vect{X}_{i}$ and $U_{i}$ are given respectively by:
\begin{align}
    \mathbb{E}[\vect{Y}_{i} | \vect{X}_{i}=\vect{x}_{i}, U_{i}=u_{i}] &= 
    \vect{\mu}_{Y} + \vect{\delta}_{Y}u_{i} + \mat{\beta}_{Y}(\vect{x}_{i}-\vect{\mu}_{X}-\vect{\delta}_{X}u_{i}), \label{eq:sn_y_conditional_mean} \\
\mathbb{E}[\vect{Z}_{i} | \vect{X}_{i}=\vect{x}_{i}, U_{i}=u_{i}] &= 
    \vect{\mu}_{Z} + \vect{\delta}_{Z}u_{i} + \mat{\beta}_{Z}(\vect{x}_{i}-\vect{\mu}_{X}-\vect{\delta}_{X}u_{i}).  \label{eq:sn_z_conditional_mean}
\end{align}
We have that $\text{var}(\vect{Y}_{i} | \vect{X}_{i}, U_{i}) = \Sigma_{YY}- {\Sigma}_{YX}{\Sigma}_{XX}^{-1}{\Sigma}_{XY}$ and  $\text{var}(\vect{Z}_{i} | \vect{X}_{i}, U_{i}) = \Sigma_{ZZ}- {\Sigma}_{ZX}{\Sigma}_{XX}^{-1}{\Sigma}_{XZ}$. The augmented likelihood for dataset A and the augmented likelihood for dataset B can be expressed as conditional regression models. Let $\vect{\alpha}_{Y} = \vect{\mu}_{Y} - \vect{\beta}_{Y} \vect{\mu}_{X},
\Omega_{Y} =  \Sigma_{YY} - \Sigma_{YX}\Sigma_{XX}^{-1}\Sigma_{XY}$ and  $\vect{\lambda}_{Y} = \vect{\delta}_{Y} - \vect{\beta}_{Y}\vect{\delta}_{X}$. Collecting terms in \eqref{eq:sn_y_conditional_mean}, the distribution of $\vect{Y}_{i}$ given $\vect{X}_{i}$ and the latent scaling variable $U_{i}$ can be represented as the regression model:
\begin{align}
\vect{Y}_{i} | \vect{X}_{i}=\vect{x}_{i}, U_{i}=u_{i} & \sim N(\vect{\alpha}_{Y} + \vect{\lambda}_{Y}u_{i} +\vect{\beta}_{Y}\vect{x}_{i}, \Omega_{Y}). \label{eq:skew_normal_y_regression}
\end{align}
Let  $\vect{\beta}_{Z} =  \Sigma_{ZX}\Sigma_{XX}^{-1},
\vect{\alpha}_{Z} = \vect{\mu}_{Z} - \vect{\beta}_{Z} \vect{\mu}_{X}, 
\Omega_{Z} =  \Sigma_{ZZ} - \Sigma_{ZX}\Sigma_{XX}^{-1}\Sigma_{XZ}$ and $\vect{\lambda}_{Z} = \vect{\delta}_{Z} - \vect{\beta}_{Z}\vect{\delta}_{X}$. Similarly, collecting terms in \eqref{eq:sn_z_conditional_mean} the conditional distribution of $\vect{Z}_{i}$ given $\vect{X}_{i}$ and the latent scaling variable $U_{i}$ can be represented as a regression model:
\begin{align}
\vect{Z}_{i} | \vect{X}_{i}=\vect{x}_{i}, U_{i}=u_{i} & \sim N(\vect{\alpha}_{Z} +\vect{\lambda}_{Z}u_{i} +\vect{\beta}_{Z}\vect{x}_{i}, \Omega_{Z}).  \label{eq:skew_normal_z_regresssion}
\end{align}
Let $\Gamma_{A}$ and $\Gamma_{B}$ contain the regression parameters for \eqref{eq:skew_normal_y_regression} and \eqref{eq:skew_normal_z_regresssion} respectively: $\Gamma_{A}=[\vect{\alpha}_{Y}\ \vect{\lambda}_{Y} \ \vect{\beta}_{Y}]^{\T}$ and $\Gamma_{B}=[\vect{\alpha}_{Z} \ \vect{\lambda}_{Z} \  \vect{\beta}_{Z}]^{\T}$. The regressions for dataset A and dataset B can be written as
\begin{align}
\mat{Y}_{A} &= \mat{B}_{A}\mat{\Gamma}_{A} + \vect{\epsilon}_{A},  \label{eq:sn_joint_y_regression_model}\\
\mat{Z}_{B} &= \mat{B}_{B}\mat{\Gamma}_{B} + \vect{\epsilon}_{B}, \label{eq:sn_joint_z_regression_model}
\end{align}
where $\vect{\epsilon}_{A} \sim MN(\vect{I}_{n_{A}}, \Omega_{Y})$ and $\vect{\epsilon}_{B} \sim MN(\vect{I}_{n_{B}}, \Omega_{Z})$. The design matrices for the regressions now include the latent $U_{i}$ terms. The complete-data design matrices are given by
\begin{align}
\mat{B}_{A} &= \begin{bmatrix}
1 & u_{1} & \vect{x}_{1}^{\mathsf{T}} \\
1 & u_{2} &\vect{x}_{2}^{\mathsf{T}} \\
\vdots \\
1 &u_{n_{A}} &\vect{x}_{n_{A}}^{\mathsf{T}}
\end{bmatrix}, 
\mat{B}_{B} = \begin{bmatrix}
1 &u_{n_{A}+1} &\vect{x}_{n_{A}+1}^{\mathsf{T}} \\
1 &u_{n_{A}+2} &\vect{x}_{n_{A}+2}^{\mathsf{T}} \\
\vdots \\
1 &u_{n_{A}+n_{B}} &\vect{x}_{n_{A}+n_{B}}^{\mathsf{T}}
\end{bmatrix}. \label{eq:design_matrix_sn}
\end{align}
The complete-data sufficient statistics are again $\mat{Y}_{A}^{\T}\mat{Y}_{A}, \mat{Z}_{B}^{\T}\mat{Z}_{B}, \mat{B}_{A}^{\T}\mat{Y}_{A}, \mat{B}_{B}^{\T}\mat{Z}_{B}, \mat{B}_{A}^{\T}\mat{B}_{A}$ and $\mat{B}_{B}^{\T}\mat{B}_{B}$.  Suppose that we impose the identification restriction  $\Sigma_{YZ} ={\Sigma}_{YX}{\Sigma}_{XX}^{-1}{\Sigma}_{XZ}$. This corresponds to a conditional independence assumption on the components of the latent random variable $\vect{V}$ in the hierarchical skew-normal model \eqref{eq:skew_normal_latent_model}. Under the the identification restriction, the  $\vect{Y}$ and $\vect{Z}$ variables are conditionally independent given $\vect{X}$ and the latent $U$. The augmented data likelihood for a single observation then has the structure
\begin{align}
f(\vect{x}_{i}, \vect{y}_{i}, \vect{z}_{i}, u_{i} ; \vect{\theta}) &= f(\vect{x}_{i},  \vect{y}_{i}, \vect{z}_{i}| u_{i}; \vect{\theta})f(u_{i}) \nonumber \\
&= f(\vect{x}_{i} | u_{i} ; \vect{\theta})f(\vect{y}_{i}|\vect{x}_{i}, u_{i}; \vect{\theta})f(\vect{z}_{i}| u_{i} ; \vect{\theta})f(u_{i}). \label{eq:skew_complete}
\end{align}
Using this property we can show the complete-data likelihood has a similar structure to the Gaussian likelihood \eqref{eq:normal_observed_likelihood}. Let $\vect{\eta}=(\vect{\eta}_{X}, \vect{\eta}_{Y}, \vect{\eta}_{Z})$ represent the parameters for each likelihood block. We have $
\vect{\eta}_{X} = (\vect{\mu}_{X}, \Sigma_{XX}, \vect{\delta}_{X})$, $
\vect{\eta}_{Y} = (\vect{\alpha}_{Y}, \vect{\beta}_{Y}, \Omega_{Y}, \vect{\lambda}_{Y})$ and $
\vect{\eta}_{Z} = (\vect{\alpha}_{Z}, \vect{\beta}_{Z}, \Omega_{Z}, \vect{\lambda}_{Z})$. 
Let $\vect{u}$ contain the latent skewing variable for each observation, so $\vect{u}=(u_{1}, \ldots, u_{n})$. Using the conditional regression models, the complete-data likelihood can be expressed as
\begin{align}
f(\vect{x}_{\text{obs}}, \vect{y}_{\text{obs}}, \vect{z}_{\text{obs}}, \vect{u} ; \vect{\eta}) = f(\vect{x}_{\text{obs}} | \vect{u}; \vect{\eta}_{X})f(\vect{y}_{\text{obs}} | \vect{x}_{\text{obs}}^{A}, \vect{u}; \vect{\eta}_{Y})f(\vect{z}_{\text{obs}}|\vect{x}_{\text{obs}}^{B}, \vect{u}; \vect{\eta}_{Z})f(\vect{u}). \label{eq:sn_complete_likelihood}
\end{align}
Equation 
\eqref{eq:sn_joint_y_regression_model} is a representation of the likelihood factor $f(\vect{y}_{\text{obs}} | \vect{x}_{\text{obs}}^{A}, \vect{u}; \vect{\eta}_{Y})$ and equation \eqref{eq:sn_joint_z_regression_model} is a representation of the likelihood factor $f(\vect{z}_{\text{obs}}|\vect{x}_{\text{obs}}^{B}, \vect{u}; \vect{\eta}_{Z})$. The E-step involves calculating the expected value of the sufficient statistics given the current parameter estimates, and can be carried out using the results in \cite{pyne2009automated}.  The M-step is simplified due to choice of parameterisation for the regression model. There are three separate maximisation tasks over $\vect{\eta}_{X}$, $\vect{\eta}_{Y}$ and $\vect{\eta}_{Z}$ that have closed form solutions. The key point is that we do not have to introduce the missing observations $\vect{y}_{\text{mis}}$ and $\vect{z}_{\text{mis}}$ to obtain a workable EM algorithm under the identification restriction $\Sigma_{YZ} ={\Sigma}_{YX}{\Sigma}_{XX}^{-1}{\Sigma}_{XZ}$. The algorithm is given in Web Appendix A. 
\subsection{Mixture of Gaussians}
\label{subsec:mixture_parametric}
We can also obtain an efficient EM algorithm for statistical matching of Gaussian mixture data using the conditional regression specifications and appropriate identification restrictions. The matching problem for Gaussian mixture data can be represented as a mixture of regression problem \citep{quandt_1972_new}. Recall the hierarchical model for the mixture model discussed in Section \ref{sec:mixture}. For independent observations $i=1, \ldots, n$:
\begin{align*}
\vect{X}_{i}, \vect{Y}_{i}, \vect{Z}_{i} | S_{i}=h &\sim f(\vect{x}_{i}, \vect{y}_{i}, \vect{z}_{i}; \vect{\theta}_{h}) \\
S_i & \sim \text{Categorical}(\pi_{1}, \ldots, \pi_{g}). 
\end{align*}
Here $f(\vect{x}_{i}, \vect{y}_{i}, \vect{z}_{i}; \vect{\theta}_{h})$ denotes a multivariate normal density with parameters $
\vect{\theta}_{h}=(\vect{\mu}^{(h)}, \Sigma^{(h)})$. The vector $\vect{\mu}^{(h)}$ gives the mean vector for component $h$ and the matrix  $\Sigma^{(h)}$ represent the covariance matrix of the $h$th component distribution for $h=1, \ldots, g$. We can partition each component mean $\vect{\mu}^{(h)}$ and each component covariance matrix $\Sigma^{(h)}$ as in \eqref{eq:gaussian_partition}. An identification restriction is then
\begin{align}
\Sigma_{YZ}^{(h)} &=  {\Sigma}_{ZX}^{(h)}{\Sigma}_{XX}^{-1 (h)}{\Sigma}_{XY}^{(h)}, \quad \text{for } h=1, \ldots,g. \label{eq:gauss_mix_component_restrictions}
\end{align}
Consider the augmented likelihood for a single observation given the latent cluster indicator $S_{i}$. Conditional on the latent cluster indicator $S_{i}$, we can again use the regression model specification from Section \ref{subsec:gaussian_matching}. Specifically,
\begin{align}
\vect{Y}_{i} | \vect{X}=\vect{x}_{i}, S_{i}=h & \sim N(\vect{\alpha}_{Y}^{(h)} +\vect{\beta}_{Y}^{(h)}\vect{x}_{i}, \Omega_{Y}^{(h)}), \label{eq:mvn_y_conditional}
\end{align}
where $\vect{\beta}_{Y}^{(h)} =  \Sigma_{YX}^{(h)}[\Sigma_{XX}^{(h)}]^{-1}$, $\vect{\alpha}_{Y}^{(h)} = \vect{\mu}_{Y}^{(h)} - \vect{\beta}_{Y}^{(h)} \vect{\mu}_{X}^{(h)}$ and $\Omega_{Y}^{(h)} =  \Sigma_{YY}^{(h)} - \Sigma_{YX}^{(h)}[\Sigma_{XX}^{(h)}]^{-1}\Sigma_{XY}^{(h)}$. The conditional distribution of $\vect{Z}_{i}$ given $\mat{X}_{i}$ obeys a similar regression equation:
\begin{align*}
\vect{Z}_{i} | \vect{X}=\vect{x}_{i}, S_{i}=h & \sim N(\vect{\alpha}_{Z}^{(h)} +\vect{\beta}_{Z}^{(h)}\vect{x}_{i}, \Omega_{Z}^{(h)}), 
\end{align*}
where $\vect{\beta}_{Z}^{(h)} =  \Sigma_{ZX}^{(h)}[\Sigma_{XX}^{(h)}]^{-1}$, $\vect{\alpha}_{Z}^{(h)} = \vect{\mu}_{Z}^{(h)} - \vect{\beta}_{Z}\vect{\mu}_{X}$ and $\Omega_{Z} =  \Sigma_{ZZ} - \Sigma_{ZX}\Sigma_{XX}^{-1}\Sigma_{XZ}$. To describe the complete-data log likelihood first let $\vect{s}$ represent the vectors of latent cluster indicators, so $\vect{s}=(s_{1}, \ldots, s_{n})$, where $s_{i} \in \left\lbrace 1, \ldots, g\right\rbrace$ for $i=1, \ldots, n$. Let $\vect{\pi}=(\pi_{1}, \ldots, \pi_{g})$ be a vector containing the mixing proportions. We can show that complete-data likelihood has a similar structure to what we obtained in for the Gaussian likelihood \eqref{eq:likelihood_factorisation}. For Gaussian mixture data, the complete-data likelihood has the structure
\begin{align}
f(\vect{x}_{\text{obs}}, \vect{y}_{\text{obs}}, \vect{z}_{\text{obs}}, \vect{s} ; \vect{\eta}, \vect{\pi}) = f(\vect{x}_{\text{obs}} | \vect{s}; \vect{\eta}_{X})f(\vect{y}_{\text{obs}} | \vect{x}_{\text{obs}}^{A}, \vect{s}; \vect{\eta}_{Y})f(\vect{z}_{\text{obs}}|\vect{x}_{\text{obs}}^{B}, \vect{s}; \vect{\eta}_{Z})f(\vect{s}; \vect{\pi}). \label{eq:mix_complete_likelihood}
\end{align}
The complete-data likelihood is the product of four components. The first likelihood block $f(\vect{x}_{\text{obs}} | \vect{s}; \vect{\eta}_{X})$ is a standard Gaussian likelihood. The second and third likelihood blocks $f(\vect{y}_{\text{obs}} | \vect{x}_{\text{obs}}^{A}, \vect{s}; \vect{\eta}_{Y})$ and $f(\vect{z}_{\text{obs}}|\vect{x}_{\text{obs}}^{B}, \vect{s}; \vect{\eta}_{Z})$ correspond to regression likelihoods. The final likelihood contribution from $f(\vect{s}; \vect{\pi})$ is the usual multinomial likelihood seen in a finite mixture. The E-step and M-step for $f(\vect{x}_{\text{obs}} | \vect{s}; \vect{\eta}_{X})$   follow from standard results on Gaussian mixture models. The E-step and M-steps for $f(\vect{y}_{\text{obs}} | \vect{x}_{\text{obs}}^{A}, \vect{s}; \vect{\eta}_{Y})$ and $f(\vect{z}_{\text{obs}}|\vect{x}_{\text{obs}}^{B}, \vect{s}; \vect{\eta}_{Z})$ follow from results on mixtures of regression models given in \cite{jones_fitting_1992}. The details are given in Web Appendix B. We avoid introducing the missing observations $\vect{y}_{\text{mis}}$ and $\vect{z}_{\text{mis}}$ into the complete-data log likelihood by using the component-wise restrictions \eqref{eq:gauss_mix_component_restrictions}. 
\subsection{Mixtures of skew-normal}
The statistical matching of data from a mixture of skew-normal distributions as a mixture of regressions problem, with latent variables in the design matrix.  For independent observations $i=1, \ldots, n$ we have that 
\begin{align*}
\vect{X}_{i}, \vect{Y}_{i}, \vect{Z}_{i} | S_{i}=h &\sim f(\vect{x}_{i}, \vect{y}_{i}, \vect{z}_{i}; \vect{\theta}_{h}) \\
S_i & \sim \text{Categorical}(\pi_{1}, \ldots, \pi_{g}). 
\end{align*}
Here $f(\vect{x}_{i}, \vect{y}_{i}, \vect{z}_{i}; \vect{\theta}_{h})$ denotes a skew-normal density with parameters $
\vect{\theta}_{h}=(\vect{\mu}^{(h)}, \Sigma^{(h)}, \vect{\delta}_{h})$. The vector $\vect{\mu}^{(h)}$ gives the mean vector for component $h$, the matrix  $\Sigma^{(h)}$ represents the scale matrix for component $h$ and $\vect{\delta}^{(h)}$ gives the skewness vector for component $h$ for $h=1, \ldots, g$. We can partition each component mean $\vect{\mu}^{(h)}$ and skewness vector $\vect{\delta}^{(h)}$ as in \eqref{eq:skew_normal_latent_model}. Each component scale matrix $\Sigma^{(h)}$ can be partitioned as in \eqref{eq:gaussian_partition}. An identification restriction is then
\begin{align}
\Sigma_{YZ}^{(h)} &=  {\Sigma}_{ZX}^{(h)}{\Sigma}_{XX}^{-1 (h)}{\Sigma}_{XY}^{(h)}, \quad \text{for } h=1, \ldots,g. \label{eq:sn_mix_restriction}
\end{align}
Introducing the latent $U_{i}$ and $S_{i}$ into the augmented likelihood, the complete-data model can be expressed in terms of conditional regression specifications. Let $\vect{\beta}_{Y}^{(h)} =  \Sigma_{YX}^{(h)}[\Sigma_{XX}^{(h)}]^{-1},
\vect{\alpha}_{Y}^{(h)} = \vect{\mu}_{Y}^{(h)} - \vect{\beta}_{Y}^{(h)} \vect{\mu}_{X}^{(h)},
\Omega_{Y}^{(h)} =  \Sigma_{YY}^{(h)} - \Sigma_{YX}^{(h)}[\Sigma_{XX}^{(h)}]^{-1}\Sigma_{XY}^{(h)}$ and  $\vect{\lambda}_{Y}^{(h)} = \vect{\delta}_{Y}^{(h)} - \vect{\beta}_{Y}^{(h)}\vect{\delta}_{X}^{(h)}$.
\begin{align}
\vect{Y}_{i} | \vect{X}_{i}=\vect{x}_{i}, U_{i}=u_{i}, S_{i}=h & \sim N(\vect{\alpha}_{Y}^{(h)} + \vect{\lambda}_{Y}^{(h)}u_{i} +\vect{\beta}_{Y}^{(h)}\vect{x}_{i}, \Omega_{Y}^{(h)}) \label{eq:mix_skew_normal_y_regression}
\end{align}
Let  $\vect{\beta}_{Z}^{(h)} =  \Sigma_{ZX}^{(h)}[\Sigma_{XX}^{(h)}]^{-1},
\vect{\alpha}_{Z}^{(h)} = \vect{\mu}_{Z}^{(h)} - \vect{\beta}_{Z}^{(h)} \vect{\mu}_{X}^{(h)}, 
\Omega_{Z}^{(h)} =  \Sigma_{ZZ}^{(h)} - \Sigma_{ZX}^{(h)}[\Sigma_{XX}^{(h)}]^{-1}\Sigma_{XZ}^{(h)}$ and $\vect{\lambda}_{Z}^{(h)} = \vect{\delta}_{Z}^{(h)} - \vect{\beta}_{Z}^{(h)}\vect{\delta}_{X}^{(h)}$. Similarly, collecting terms in \eqref{eq:sn_z_conditional_mean} the conditional distribution of $\vect{Z}_{i}$ given $\vect{X}_{i}$ and the latent scaling variable $U_{i}$ can be represented as a regression model
\begin{align}
\vect{Z}_{i} | \vect{X}_{i}=\vect{x}_{i}, U_{i}=u_{i}, S_{i}=h & \sim N(\vect{\alpha}_{Z}^{(h)} +\vect{\lambda}_{Z}^{(h)}u_{i} +\vect{\beta}_{Z}^{(h)}\vect{x}_{i}, \Omega_{Z}^{(h)}).  \label{eq:mix_skew_normal_z_regresssion}
\end{align}
The complete-data likelihood has the structure
\begin{align}
f(\vect{x}_{\text{obs}}, \vect{y}_{\text{obs}}, \vect{z}_{\text{obs}}, \vect{s}, \vect{u} ; \vect{\eta}, \vect{\pi}) = f(\vect{x}_{\text{obs}} | \vect{s}, \vect{u}; \vect{\eta}_{X})f(\vect{y}_{\text{obs}} | \vect{x}_{\text{obs}}^{A}, \vect{s}, \vect{u}; \vect{\eta}_{Y})f(\vect{z}_{\text{obs}}|\vect{x}_{\text{obs}}^{B}, \vect{s}, \vect{u}; \vect{\eta}_{Z})f(\vect{s}; \vect{\pi})f(\vect{u}) \label{eq:sn_mix_complete_likelihood}
\end{align}
The first likelihood block $f(\vect{x}_{\text{obs}} | \vect{s}, \vect{u}; \vect{\eta}_{X})$ corresponds to a skew-normal likelihood. The second and third likelihood blocks are regression likelihoods with the latent $u$ variables in the design matrices. We can again define an EM algorithm for parameter estimation under the identification constraint \eqref{eq:sn_mix_restriction}. The E-step is carried out using existing results for the skew-normal distribution. The M-step is again simplified by the choice of parameterisation. We have three separate maximisation tasks over $\vect{\eta}_{X}$, $\vect{\eta}_{Y}$ and $\vect{\eta}_{Z}$. The algorithm is given in full in Web Appendix C. Once again we do not introduce the missing observations $\vect{y}_{\text{mis}}$ and $\vect{z}_{\text{mis}}$ into complete-data likelihood by using the identification restriction \eqref{eq:sn_mix_restriction}.

\section{Nearest-neighbour matching}
Nearest-neighbour imputation is a popular alternative to parametric methods in data fusion problems \citep{aluja2007graft, saporta_2002_data}. Nearest-neighbour methods rely on the conditional independence assumption, that is $\vect{Y}$ and $\vect{Z}$ are conditionally independent given $\vect{X}$. The nearest-neighbour method matches observations in dataset A and dataset B based on the Euclidean distance measured using the common $\vect{X}$ dimensions. Missing values are imputed by taking values from the nearest-neighbour in the donor set. More formally, the missing  $\vect{Z}$ values for observation $i$ in dataset $A$ are set as
\begin{align*}
\vect{z}_{i} = \vect{z}_{k}, \quad  \text{ where }k = \underset{k \in \set{n_{A}+1,\ \ldots , n_{A}+n_B}}{\operatorname{argmin}}  ||\vect{x}_{i}-\vect{x}_{k}||_{2},
\end{align*}
for $i=1, \ldots, n_{A}$. Similarly, the missing $\vect{Y}$ values  for observation $i$ in dataset B are set as 
\begin{align*}
\vect{y}_{i}= \vect{y}_{k}, \quad  \text{ where }k = \underset{k \in \set{1,\ \ldots , n_A}}{\operatorname{argmin}}  ||\vect{x}_{i}-\vect{x}_{k}||_{2},
\end{align*}
for $i=n_{A}+1, \ldots, n_{A}+n_{B}$. As discussed by \cite{rassler_2002_statistical},  each of the observations in imputed dataset can be viewed as exchangeable draws from some distribution $g(\vect{x}_{i}, \vect{y}_{i}, \vect{z}_{i})$. The distribution $g(\vect{x}_{i}, \vect{y}_{i}, \vect{z}_{i})$ is useful to characterise the behaviour of the nearest-neighbour method. \cite{marella2008matching} consider the asymptotic form of the nearest-neighbour imputation distribution and show that as the size of the donor set tends to infinity, 
\begin{align}
g(\vect{x}_{i}, \vect{y}_{i}, \vect{z}_{i}) \to f(\vect{x}_{i}; \vect{\theta})f(\vect{y}_{i}|\vect{x}_{i}; \vect{\theta})f(\vect{z}_{i}|\vect{x}_{i}; \vect{\theta}). \label{eq:nn_asymptotic}
\end{align}
In  equation \eqref{eq:nn_asymptotic} $f(\vect{x}_{i}; \vect{\theta}),f(\vect{y}_{i}|\vect{x}_{i}; \vect{\theta})$ and $f(\vect{z}_{i}|\vect{x}_{i}; \vect{\theta})$ are the marginal and conditional distributions from the true generative model $f(\vect{x}_{i},\vect{y}_{i},\vect{z}_{i}; \vect{\theta})$.
The nearest-neighbour strategy will only produce the correct joint distribution if $\vect{Y}$ and $\vect{Z}$ are conditionally independent given $\vect{X}$. For multivariate-normal data, the asymptotic form of the nearest-neighbour imputation scheme is
\begin{align}
 g(\vect{x}_{i}, \vect{y}_{i}, \vect{z}_{i}) &=N_{p}\left(
\vect{\mu} = \begin{bmatrix} \bm{\mu}_X \\ \bm{\mu}_Y \\ \bm{\mu}_Z \end{bmatrix}, {\Sigma} = 
\begin{bmatrix}
{\Sigma}_{XX} & {\Sigma}_{XY} & {\Sigma}_{XZ} \\
{\Sigma}_{YX} & {\Sigma}_{YY} & {\Sigma}_{YX}{\Sigma}_{XX}^{-1}{\Sigma}_{XZ} \\
{\Sigma}_{ZX} & \Sigma_{ZX}\Sigma_{XX}^{-1}\Sigma_{YX}  &{\Sigma}_{ZZ}
\end{bmatrix} \right).
\end{align}
The distribution  $g(\vect{x}_{i}, \vect{y}_{i}, \vect{z}_{i})$ is equivalent to the distribution recovered by the maximum likelihood parametric approach under the identification constraint $\Sigma_{YZ} = {\Sigma}_{YX}{\Sigma}_{XX}^{-1}{\Sigma}_{XZ}$. The asymptotic equivalence between nearest-neighbour matching and a parametric methods may not necessarily hold for non-Gaussian data.  

It can be difficult to justify the conditional independence assumption for non-Gaussian data as there may be no parameter $\vect{\theta}^{*}$ such that $f(\vect{x}, \vect{y}, \vect{z}; \vect{\theta}^{*}) = f(\vect{x}; \vect{\theta}^{*})f(\vect{y}| \vect{x};\vect{\theta}^{*})f(\vect{z} | \vect{x} ; \vect{\theta}^{*})$. This problem is likely to arise when there is some latent structure in the generative model that links the $\vect{Y}$ and $\vect{Z}$ variables. In these situations, the nearest-neighbour imputation distribution
$g(\vect{x}_{i}, \vect{y}_{i}, \vect{z}_{i}) = f(\vect{x} ; \vect{\theta}) f(\vect{y}|\vect{x}_{i} ; \vect{\theta})f(\vect{z}|\vect{x}; \vect{\theta})$ may not be a good approximation for the true generative model $f(\vect{x}, \vect{y}, \vect{z}; \vect{\theta})$. Even in situations where the generative model is identifiable, the nearest-neighbour scheme may fail to produce statistically sound imputations. We examine this issue for skew-normal distribution and Gaussian mixtures.

\subsection{Skew-normal}
\label{sec:skew_normal}
\begin{figure}
    \centering
    \includegraphics[width=\textwidth]{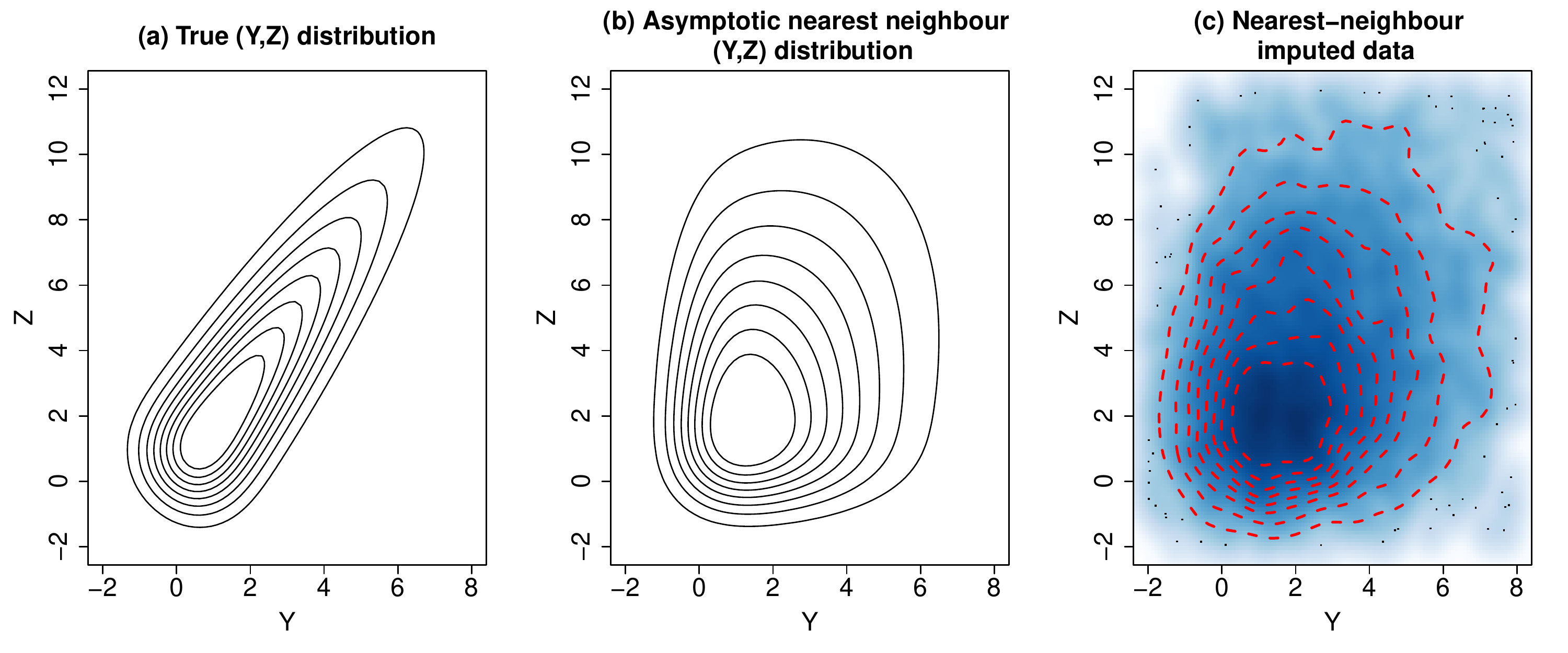}
    \caption{Analysis of nearest-neighbour imputation for skewed data. Panel (a) shows contours of the true $(Y, Z)$ joint distribution. Panel (b) shows contours of the nearest-neighbour imputation distribution $g(\vect{y}, \vect{z})$. Panel (c) gives a smoothed scatter plot of the imputed data in the simulation. The red-dashed lines given contours from a kernel density estimate using the imputed data. Nearest-neighbour imputation underestimates the $\vect{Y}, \vect{Z}$ correlation.}
    \label{fig:sn_theoretical}
\end{figure}

In general, the imputation of skewed data can be challenging relative to the multivariate normal case \citep{sterne_2009_multiple}, and the statistical matching of skew-normal data presents some difficulties.  
Assuming that all variables have some skewness, so all elements of $\vect{\delta}$ are nonzero, it is not possible for $\vect{Y}$ and $\vect{Z}$ to be conditionally independent given $\vect{X}$ \citep{azzalini_1999_statistical}. Contrary to the normal case, there is no parameter constraint on $\vect{\mu}, \Sigma, \vect{\delta}$ such that $
f(\vect{x}_{i}, \vect{y}_{i}, \vect{z}_{i}; \vect{\mu}, \Sigma, \vect{\delta}) = f(\vect{x}_{i}; \vect{\mu}, \Sigma, \vect{\delta})f(\vect{y}_{i}| \vect{x}_{i}; \vect{\mu}, \Sigma, \vect{\delta})f(\vect{z}_{i}| \vect{x}_{i}; \vect{\mu}, \Sigma, \vect{\delta})$. The absence of a conditional independence constraint poses problems for nearest-neighbour imputation. To illustrate, we generated data from a three-dimensional skew-normal distribution. The  parameters were set as $\vect{\mu}=\vect{0}$, $\Sigma=\mat{I}$ and $\vect{\delta}=(1,3,5)^{\T}$. 
The first, second and third dimensions were labelled as $\vect{X}$, $\vect{Y}$ and $\vect{Z}$ respectively. We generated two datasets with $n_{A}=n_{B}=5000$. Panels (a) and (b) in Figure \ref{fig:sn_theoretical} compare the true joint $(\vect{Y}, \vect{Z})$ distribution to the nearest-neighbour asymptotic distribution. The nearest-neighbour distribution shows a much weaker linear association between the $\vect{Y}$ and $\vect{Z}$ variables compared to the true distribution. The contours in panel (b) are more square shaped than the skewed ellipse in panel (a). We applied nearest-neighbour imputation to the simulated dataset to check the correspondence with the theoretical distribution in (b).  Panel (c) shows a smoothed scatter plot of the imputed data along with contours from a kernel density estimate as red dashed lines. The contours of the density estimate closely resemble the theoretical contours in (b), and are again more square shaped than the contours of the true $(\vect{Y}, \vect{Z})$ distribution in (a). 

\subsection{Finite mixture models}
\label{sec:mixture}
Finite mixture models are well suited for the statistical analysis of flow cytometry data. To describe the model, let $\vect{\Psi}$ contain the mixing proportions $\pi_{1}, \ldots, \pi_{g}$ and parameters $\vect{\theta}_{1}, \ldots, \vect{\theta}_{g}$ for each component. The $j$th component distribution can be represented as $f(\vect{x}, \vect{y}, \vect{z}; \vect{\theta}_{j})$ for $j=1, \ldots, g$. The distribution function of the $g$-component mixture model is given by
\begin{align}
f(\vect{x}_{i}, \vect{y}_{i}, \vect{z}_{i}; \vect{\Psi}) &= \sum_{j=1}^{g}\pi_{j}f(\vect{x}_{i}, \vect{y}_{i}, \vect{z}_{i}; \vect{\theta}_{j}).
\end{align}
It will be helpful to conceptualise the mixture model in terms of  a latent cluster indicator $S_{i} \sim \text{Categorical}(\pi_{1}, \ldots, \pi_{g})$ for each observation $i=1, \ldots, n$. The generative model has the hierarchical representation:
\begin{align*}
\vect{X}_{i}, \vect{Y}_{i}, \vect{Z}_{i} | S_{i} = h &\sim f(\vect{x}_{i}, \vect{y}_{i}, \vect{z}_{i}; \vect{\theta}_{h}) \\
S_{i} & \sim \text{Categorical}(\pi_{1}, \ldots, \pi_{g}). 
\end{align*}

It is unlikely for the conditional independence assumption to be appropriate when the generative model is a finite mixture. Given a mixture model with well-separated components, it is not possible to find an appropriate restrictions on the mixture parameters $\vect{\Psi}$ such that $
f(\vect{x}_{i}, \vect{y}_{i}, \vect{z}_{i}; \vect{\Psi}) = f(\vect{x}_{i}; \vect{\Psi})f(\vect{y}_{i}| \vect{x}_{i}; \vect{\Psi})f(\vect{z}_{i}| \vect{x}_{i}; \vect{\Psi})$. Conditional on $\vect{X}$, the remaining $\vect{Y}$ and $\vect{Z}$ variables are almost certainly dependent due to the latent cluster indicator $S$. The violation of the conditional independence assumption for finite mixture models means that the nearest-neighbour imputation may have undesirable behaviour.  The tendency for nearest-neighbour imputation to produce spurious clusters has been demonstrated empirically in statistical matching problems in flow cytometry analysis \citep{lee2011statistical, oneill_2015_deep}. 

From the analysis of the asymptotic model, spurious clusters emerge when the nearest-neighbour match originates from a different mixture component to the query point $\vect{x}_{i}$. The probability of observing spurious clusters is related to how informative the common $\vect{X}$ variables are for classification. If the clusters are poorly separated using the $\vect{X}$ variables, then there is a high probability of mismatching observations across clusters and giving improper imputations.

\begin{figure}
    \centering
    \includegraphics[width=\textwidth]{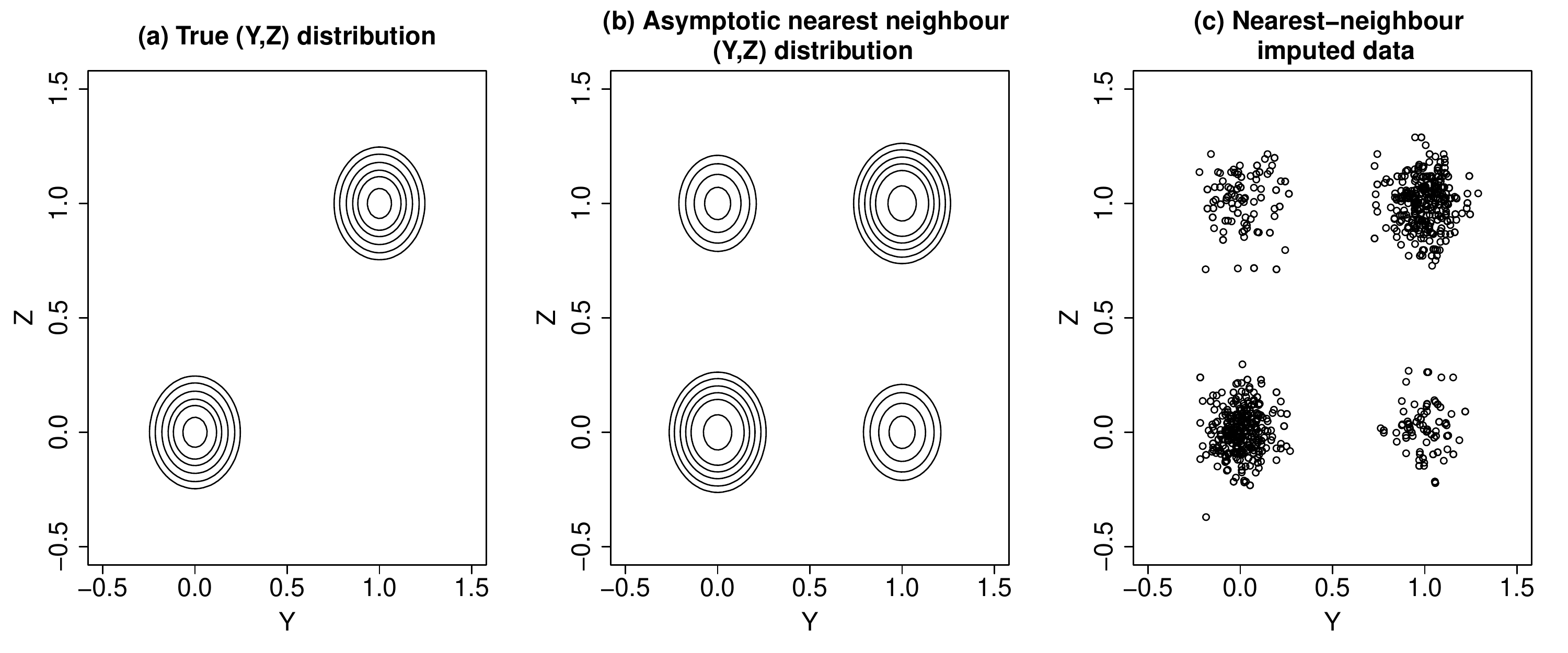}
    \caption{Analysis of nearest-neighbour imputation for Gaussian mixture data. Panel (a) shows contours of the true $(Y, Z)$ joint distribution. Panel (b) shows contours of the nearest-neighbour imputation distribution $g(\vect{y}, \vect{z})$. Panel (c) gives a shows the imputed data from the simulation. Nearest-neighbour imputation introduces spurious clusters.}
    \label{fig:gaussian_mixture_theoretical}
\end{figure}

To illustrate, we generated data from a equally weighted two component Gaussian mixture model. The first, second and third dimensions were labelled as the $\vect{X}$, $\vect{Y}$ and $\vect{Z}$ variables respectively. Component 1 had a mean of $(-0.1, 0, 0)^{\mathsf{T}}$ and component 2 had a mean of $(0.1, 1,1)^{\mathsf{T}}$. Each component had covariance matrix $0.01\mat{I}$. The components are well separated in the marginal $\vect{Y}$ and $\vect{Z}$ dimensions and there is a large amount of overlap in the $\vect{X}$ dimension. We generated $n_{A}=n_{B}=500$ observations in each dataset. Panels (a) and (b) in Figure \ref{fig:gaussian_mixture_theoretical} compare the true distribution  on the $\vect{Y}$ and $\vect{Z}$ variables to the asymptotic nearest neighbour distribution. The nearest-neighbour distribution shows four clusters in the joint distribution when there should only be two. This is because the $\vect{X}$ variables contain limited information for clustering. Panel (c) shows the results of applying nearest-neighbour matching to the simulated dataset. The introduction of two spurious clusters is consistent with the predicted behaviour from panel (b).

\section{Examples}
\subsection{Skew-normal}
\begin{figure}[h]
\centering
\includegraphics[width=\textwidth]{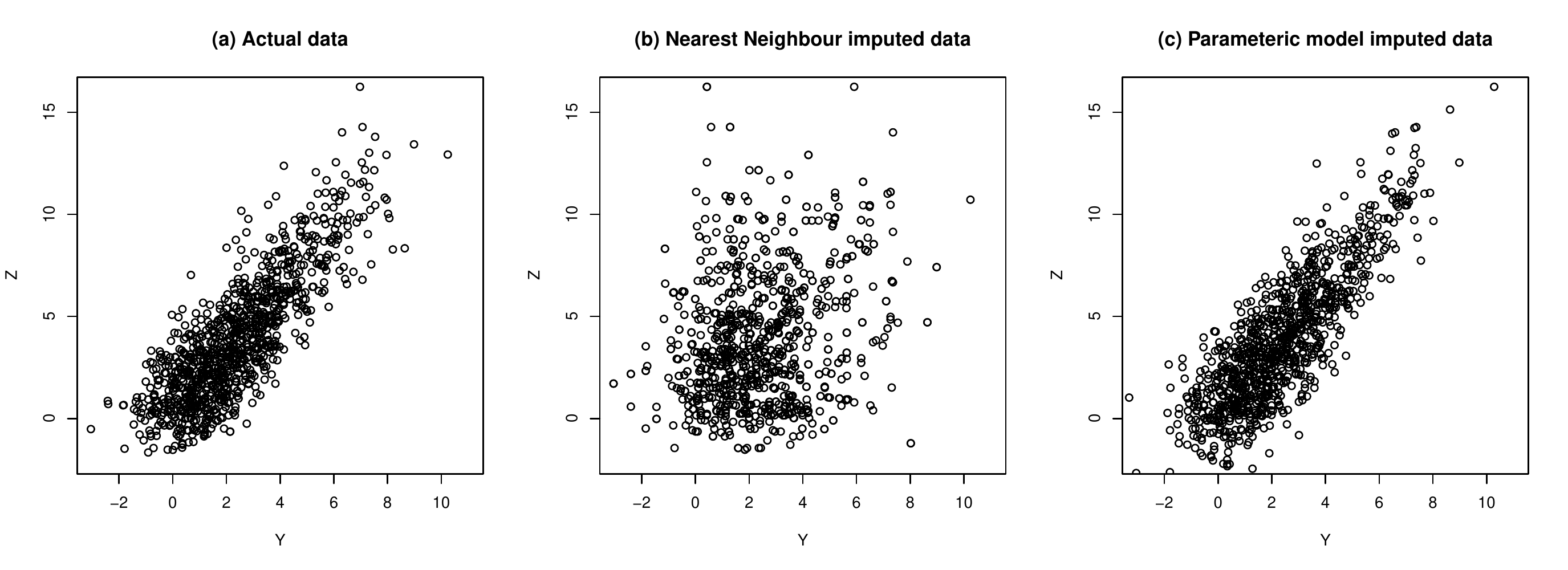}
\caption{Comparison of imputation strategies on skew-normal data.}
\label{fig:skew_normal_nn}
\end{figure}

We return to the skew-normal example in Section \ref{sec:skew_normal}. We simulated another two datasets with $n_{A}=n_{B}=500$. The parameters were again set as $\vect{\mu}=\vect{0}$, $\Sigma=\mat{I}$ and $\vect{\delta}=(1,3,5)^{\T}$. 
The first, second and third dimensions were labelled as $\vect{X}$, $\vect{Y}$ and $\vect{Z}$ respectively. We applied nearest-neighbour matching and parametric imputation using the constraint $\Sigma_{YZ}={\Sigma}_{YX}{\Sigma}_{XX}^{-1}{\Sigma}_{XZ}$. Parameters were estimated using maximum likelihood. The results are shown in Figure \ref{fig:skew_normal_nn}. In this example, there is a large difference between the nearest-neighbour imputations (b) and the parametric imputations (c). The true sample correlation between the $Y$ and $Z$ variables is $\rho_{YZ} = 0.83$.  Nearest-neighbour imputation gives an underestimate, with $\widehat{\rho}_{YZ}=0.14$. Parametric imputation gives much better estimate $\widehat{\rho}_{YZ}=0.85$. The $(Y,Z)$ distribution is recoverable in this scenario, as the generative model satisfies the identification restriction  $\Sigma_{YZ}={\Sigma}_{YX}{\Sigma}_{XX}^{-1}{\Sigma}_{XZ}$. 

\section{Data application}
\subsection{Iris dataset}
\begin{figure}
    \centering
    \includegraphics[width=\textwidth]{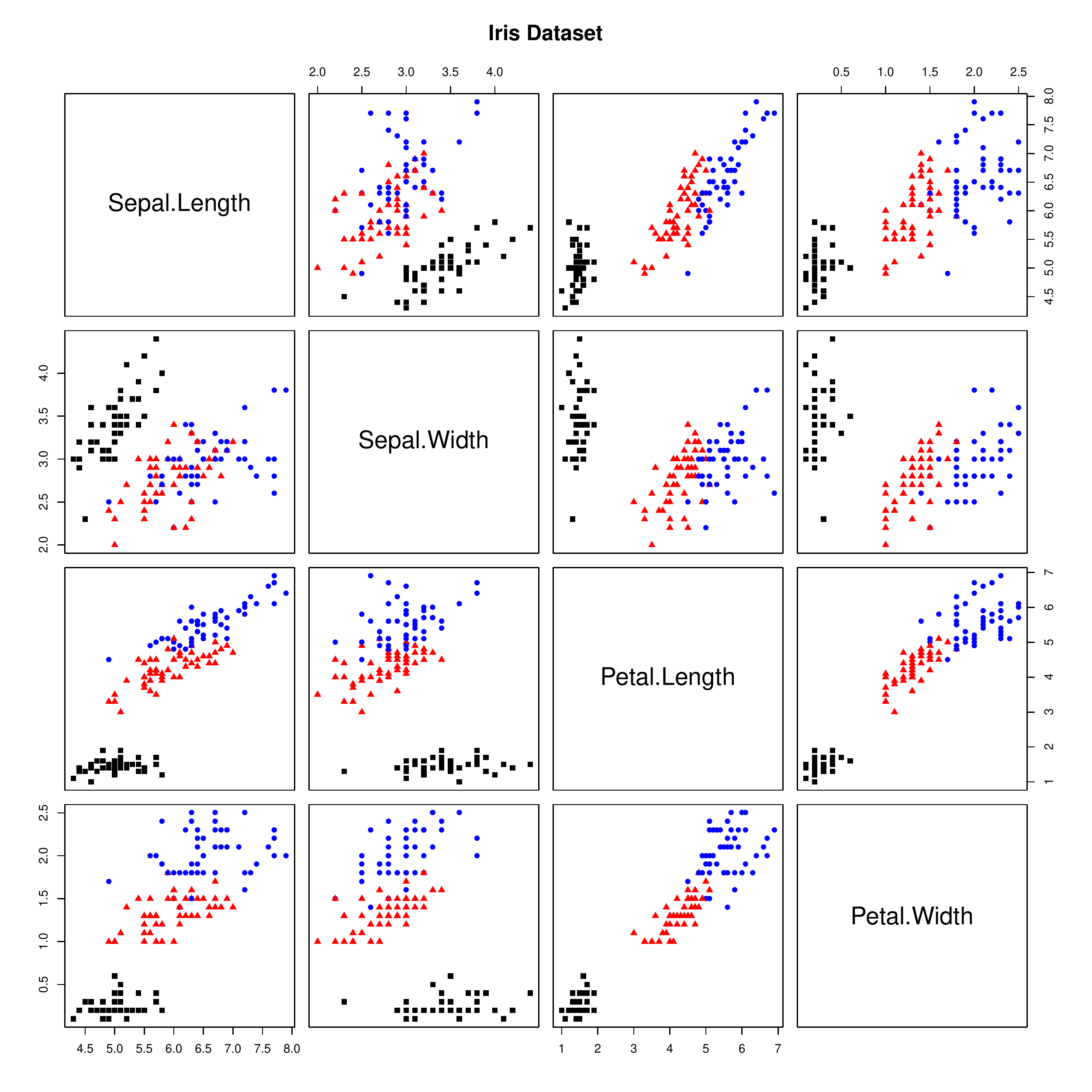}
    \caption{Iris dataset. Black squares, red triangles and blue cirles denote observations from the species `setosa',  `virginica' and `versicolor' repsectively.}
    \label{fig:iris_pairs}
\end{figure}

We analysed a Anderson's iris dataset \citep{anderson_iris_1935,fisher_use_1936} using a Gaussian mixture model. Figure \ref{fig:iris_pairs} shows a pairs plot of the dataset with colour and shape giving the species labels. 

We considered two matching scenarios, under the assumption that the species labels were unavailable. These are described in Table \ref{tab:iris_matching_scenarios}. Nearest-neighbour matching was compared to parametric imputation using the same identification constraints in Section \ref{sec:mixture}. An important difference between the two scenarios is that in Scenario 1, the $\vect{X}$ single variable sepal with does not give enough information to reliably classify the observations into species. In Scenario 2, the sepal length and sepal width together give enough information to reliably classify the observations into species.  

\begin{figure}[h]
\centering
\includegraphics[width=\textwidth]{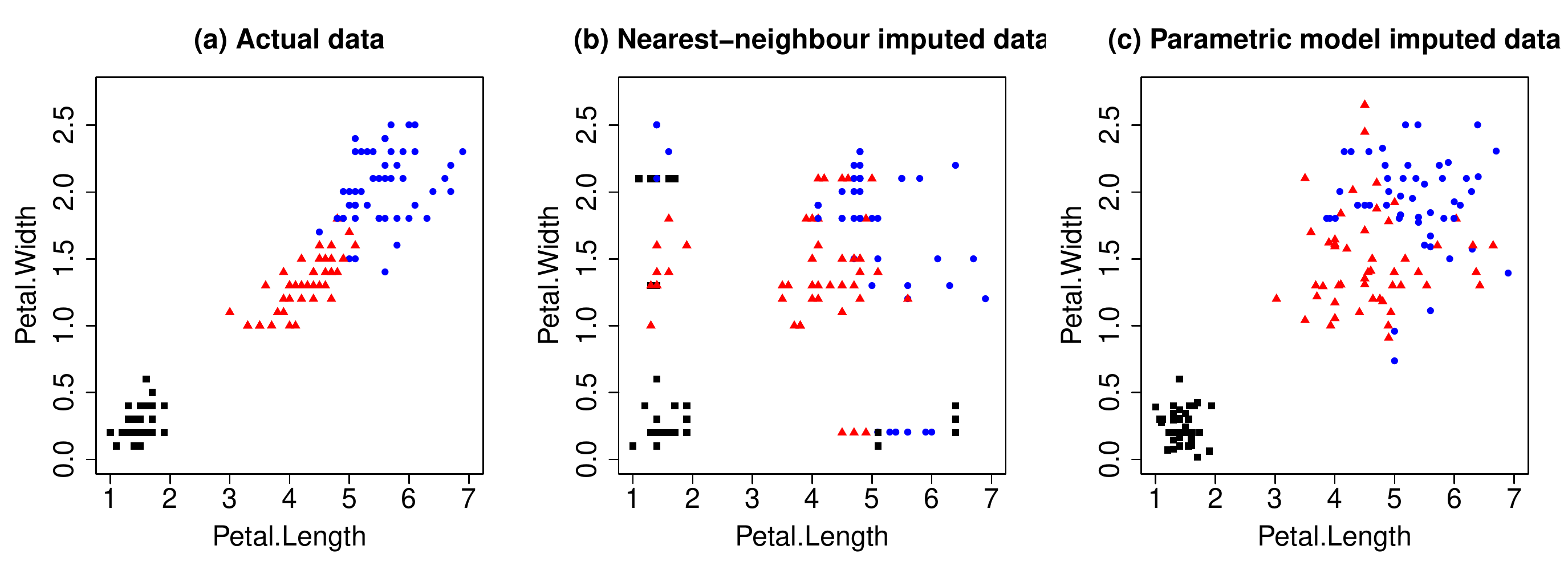}
\caption{Actual and imputed data in Scenario 1 for the iris dataset.}
\label{fig:iris_example_1}
\end{figure}

Figure \ref{fig:iris_example_1} shows the results for Scenario 1. The asymptotic analysis showed that when the $\vect{X}$ variables are uninformative for clustering, we expect nearest-neighbour imputation to produce spurious clusters. In this example, we encounter mismatched species labels across datasets A and B when using sepal width to find nearest-neighbours. The parametric strategy does not create spurious clusters. The restrictions on the model based clustering approach encode a different set of assumptions that are more reasonable for a Gaussian mixture. As such, the parametric model imputed data shows more fidelity with the true data. 

\begin{table}[h]
\centering

\begin{tabular}{llll}
\toprule
&$\vect{X}$  & $\vect{Y}$ & $\vect{Z}$  \\ 
\midrule
Scenario 1 \qquad  & Sepal Width \qquad  \quad & Petal Length  \qquad  \quad & Petal Width  \\  
Scenario 2 \qquad  & Sepal Width, Sepal Length \qquad  \quad & Petal Length  \qquad  \quad & Petal Width  \\  
\bottomrule                 
\end{tabular}

\caption{Matching scenarios using the iris dataset. Clustering using the $\vect{X}$ variables is substantially easier in Scenario 2.}
\label{tab:iris_matching_scenarios}
\end{table}

\begin{figure}[h]
\centering
\includegraphics[width=\textwidth]{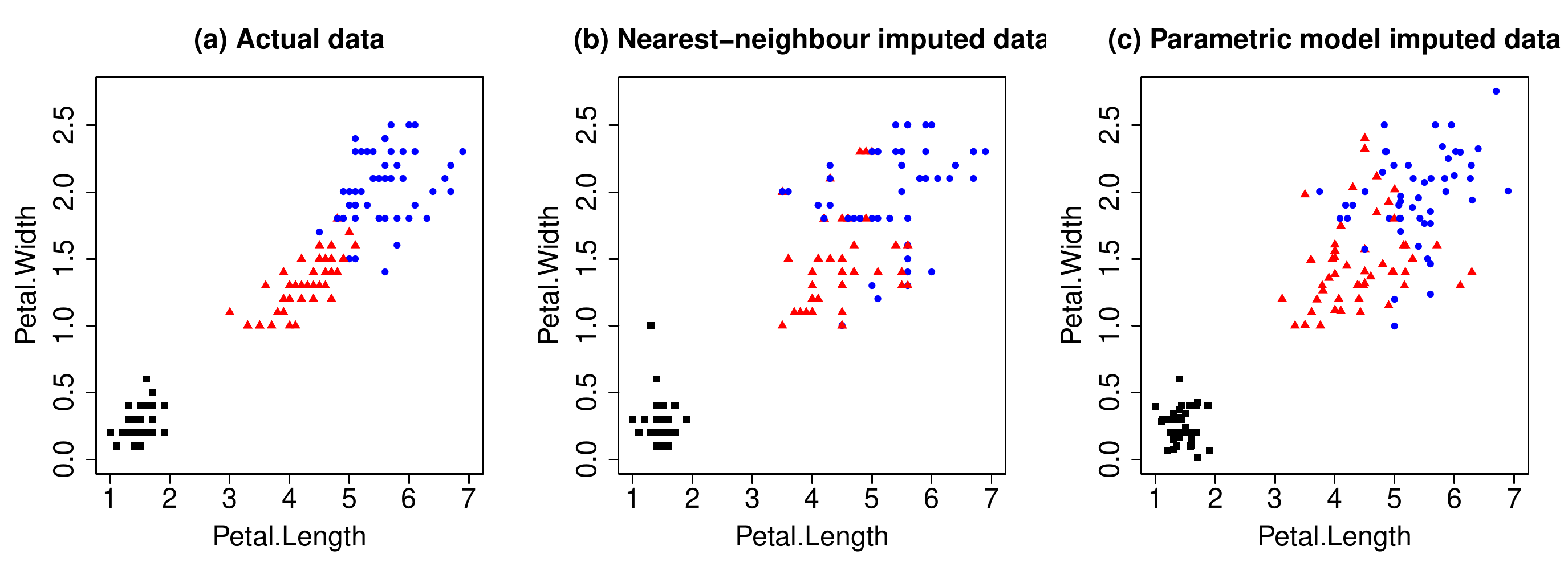}
\caption{Actual and imputed data in Scenario 2 for the iris dataset.}
\label{fig:iris_example_2}
\end{figure}

Figure \ref{fig:iris_example_2} shows the results for Scenario 2. In this case nearest-neighbour imputation does not produce any spurious clusters. Spurious clusters are not expected here as the addition of sepal length into the common $\vect{X}$ group gives enough information to cluster the observations. The additional information should greatly reduce the number of mismatched species labels when finding nearest-neighbours across datasets. Parametric imputation gives very similar results to the nearest-neighbour method in Scenario 2. In both scenarios the imputed data shows weaker correlations than in the true dataset, this is because the component-wise identification restriction $\Sigma_{YZ}^{(h)}={\Sigma}_{YX}^{(h)}[{\Sigma}_{XX}^{(h)}]^{-1}{\Sigma}_{XZ}^{(h)}$ for $h=1, \ldots, g$  is perhaps not appropriate for this dataset.

\subsection{Flow cytometry data (single cell subpopulation)}
\begin{figure}[h]
\centering
\includegraphics[width=0.67\textwidth]{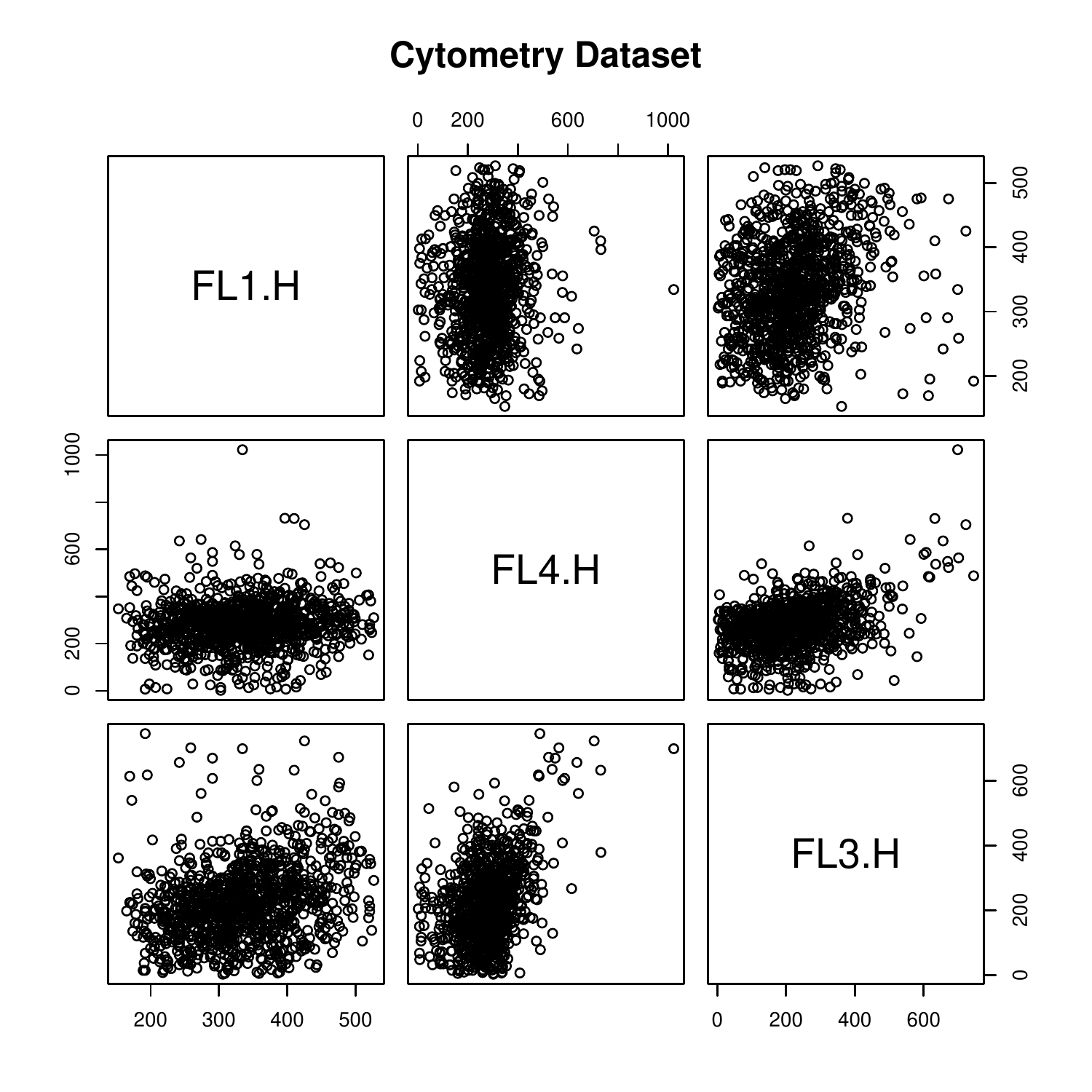}
\caption{Flow cytometry dataset with a single cell subpopulation.}
\label{fig:cytometry_example}
\end{figure}

\begin{figure}[h]
\centering
\includegraphics[width=\textwidth]{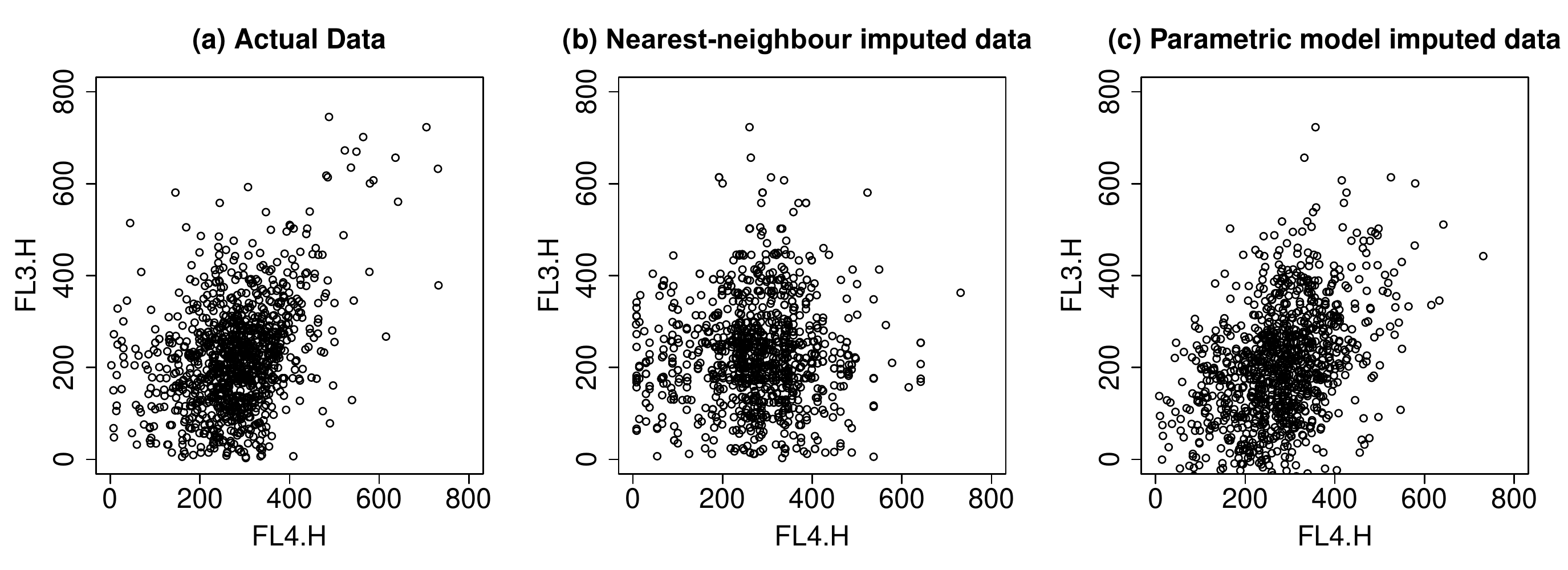}
\caption{Actual and imputed data for the single cell population cytometry dataset}
\label{fig:cytometry}
\end{figure}

We analysed a subset of flow cytometry data from a study on graft vs host disease \citep{brinkman_2007_high}. See the supplementary material for more information on the dataset. The dataset is displayed in Figure \ref{fig:cytometry_example}. The data subset has $n=1162$ observations on $p=3$ markers. The dataset was split into two datasets of $n_{A}=n_{B}=681$ observations. We took FL1.H, FL4.H and FL3.H as the $\vect{X}$, $\vect{Y}$ and $\vect{Z}$ variables respectively. We applied nearest-neighbour matching and parametric imputation with a skew-normal model using the identification restriction described in Section \ref{subsec:sn_parametric}. Figure \ref{fig:cytometry} compares the imputed data to the actual data. Nearest-neighbour imputation gives different results than the parametric approach. The nearest-neighbour imputed data seems to exhibit the broad shape of the asymptotic imputation distribution that was studied in Section \ref{sec:skew_normal}. If the skew-normal model is appropriate we expect to see weaker joint associations between the $\vect{Y}$ and $\vect{Z}$ variables in the imputed dataset than in the true dataset. It appears that the parametric model does a better job of recovering the general shape of the $(\vect{Y}, \vect{Z})$ joint distribution. The true sample correlation between the $(\vect{Y}, \vect{Z})$ variables is  $\rho_{YZ}=0.39$. Parametric imputation gives a better estimate of the correlation ($\widehat{\rho}_{YZ}=0.44$) compared to nearest-neighbour imputation ($\widehat{\rho}_{YZ} = 0.03$). 

\subsection{Flow cytometry data (two cell subpopulations)}
\begin{figure}[h]
\centering
\includegraphics[width=0.67\textwidth]{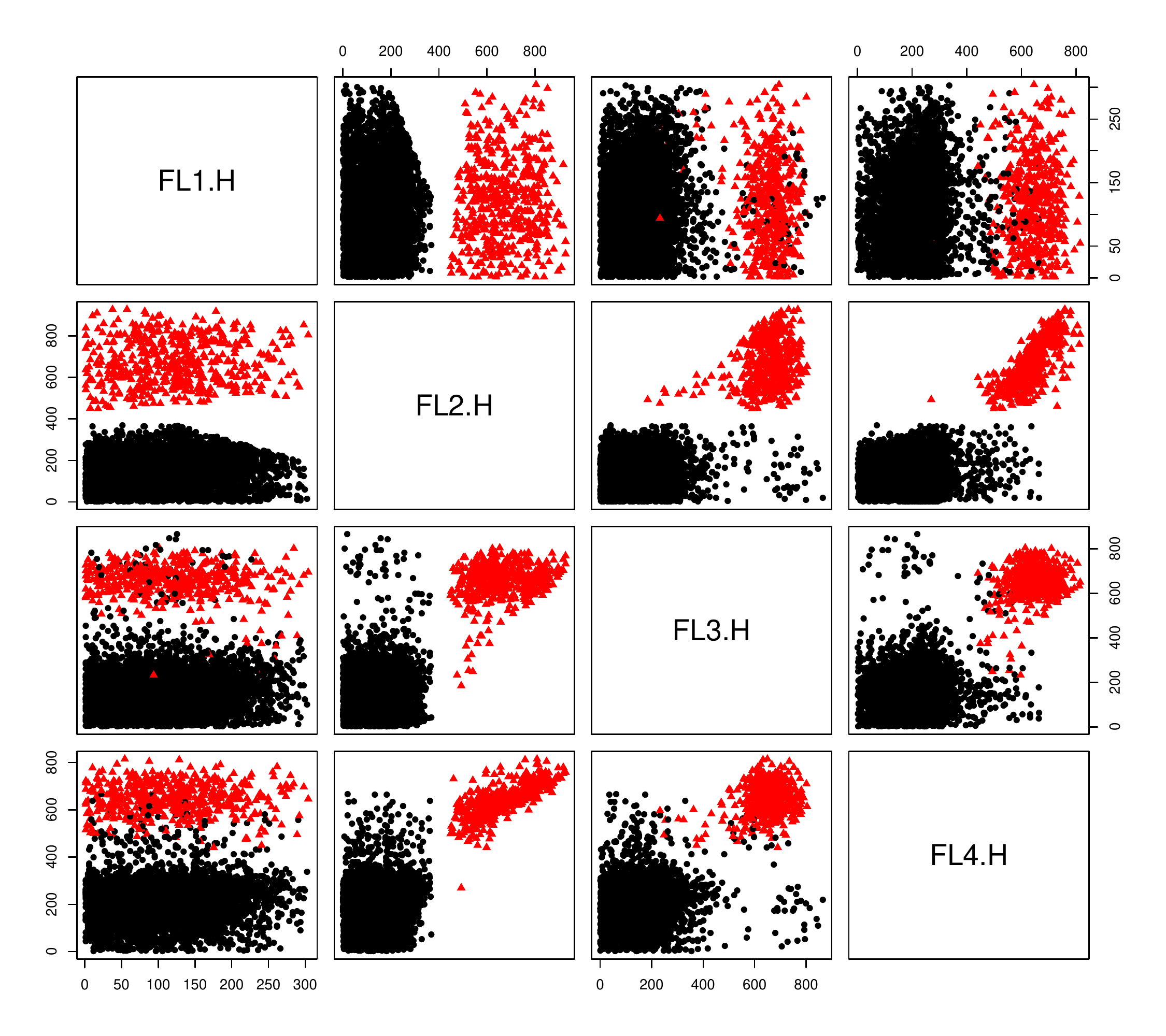}
\caption{Two cell subpopulation data. Black circles and red triangles denote observations from cell subpopulations 1 and 2 respectively. }
\label{fig:cytometry_mixture_dataset}
\end{figure}

We also consider another dataset from the graft vs host study. We extracted data on two cell subpopulations that were not considered in the previous data example. The data subset has $n=1321$ observations on $p=3$ markers. The dataset was split into two datasets of $n_{A}=660$, $n_{B}=661$ observations, and is plotted in Figure \ref{fig:cytometry_mixture_dataset}. We considered two different matching scenarios described in Table \ref{tab:cytometry_matching}. 

\begin{table}
\centering
\begin{tabular}{llll}
\toprule
&$\vect{X}$  & $\vect{Y}$ & $\vect{Z}$  \\ 
\midrule
Scenario 1 \qquad  & FL1.H \qquad  \quad & FL4.H  \qquad  \quad & FL2.H \\
Scenario 2 \qquad  & FL3.H \qquad  \quad & FL4.H  \qquad  \quad & FL2.H \\   
\bottomrule
\end{tabular}
\label{tab:cytometry_matching}
\caption{Matching scenarios using the two cell subpopulation flow cytometry dataset. Clustering using the $\vect{X}$ variables is substantially easier in Scenario 2.}
\end{table}

\begin{figure}[h]
\centering
\includegraphics[width=\textwidth]{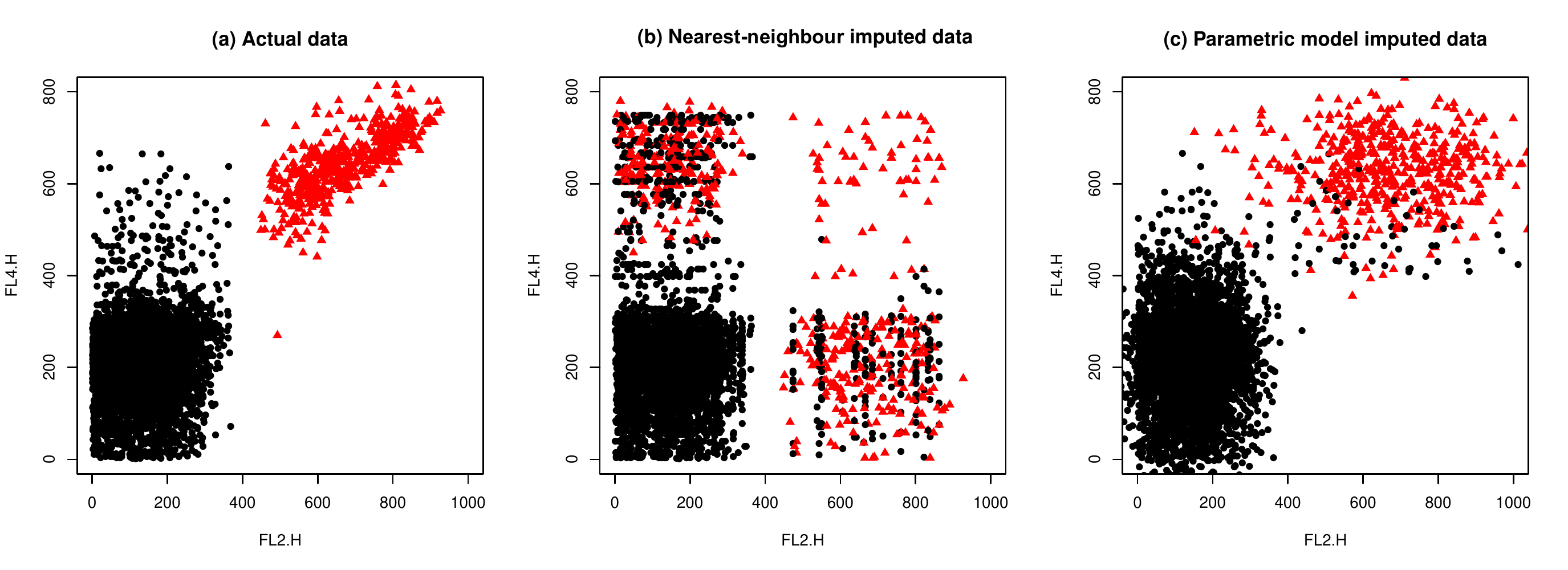}
\caption{Actual and imputed data for the two cell population cytometry dataset in Scenario 1. Black circles and red triangles denote observations from cell subpopulations 1 and 2 respectively. }
\label{fig:cytometry_mixture_scenario1}
\end{figure}

We applied nearest-neighbour matching and parametric imputation with a skew-normal mixture model in each scenario. Figure \ref{fig:cytometry_mixture_scenario1} compares the imputed data to the actual and data in Scenario one. The matching variable FL3.H is not sufficiently informative for clustering in Scenario 1, and nearest-neighbour imputation introduces spurious clusters. Focusing on panel (b), we see that the majority of points from the second cell subpopulation (red diamonds) are no longer located in the top right corner of the plot. Nearest-neighbour matching shifts most observations in cell subpopulation 2 to either the top-left or bottom-right of the plot. This is a serious distortion of the joint relationship between the FL4.H and FL2.H variables that is present in the original dataset. Looking at panel (c) we see that parameteric imputation does not suffer from this problem. We recover the correct general location of each cell cluster in the joint space of the FL4.H and FL2.H variables. Although the location of the second cell subpopulation is recovered well, we do not preserve the correct orientation of the second cell subpopulation.

Table \ref{tab:cytometry_correlations} reports sample correlations for each cell subpopulation in Scenario one.  Nearest-neighbour imputation gives poor estimates of the subpopulation correlations due to the mismatching of observations across clusters. In this scenario the parametric imputation strategy underestimates the $\vect{Y}, \vect{Z}$ correlation for both groups.

\begin{figure}[h]
\centering
\includegraphics[width=\textwidth]{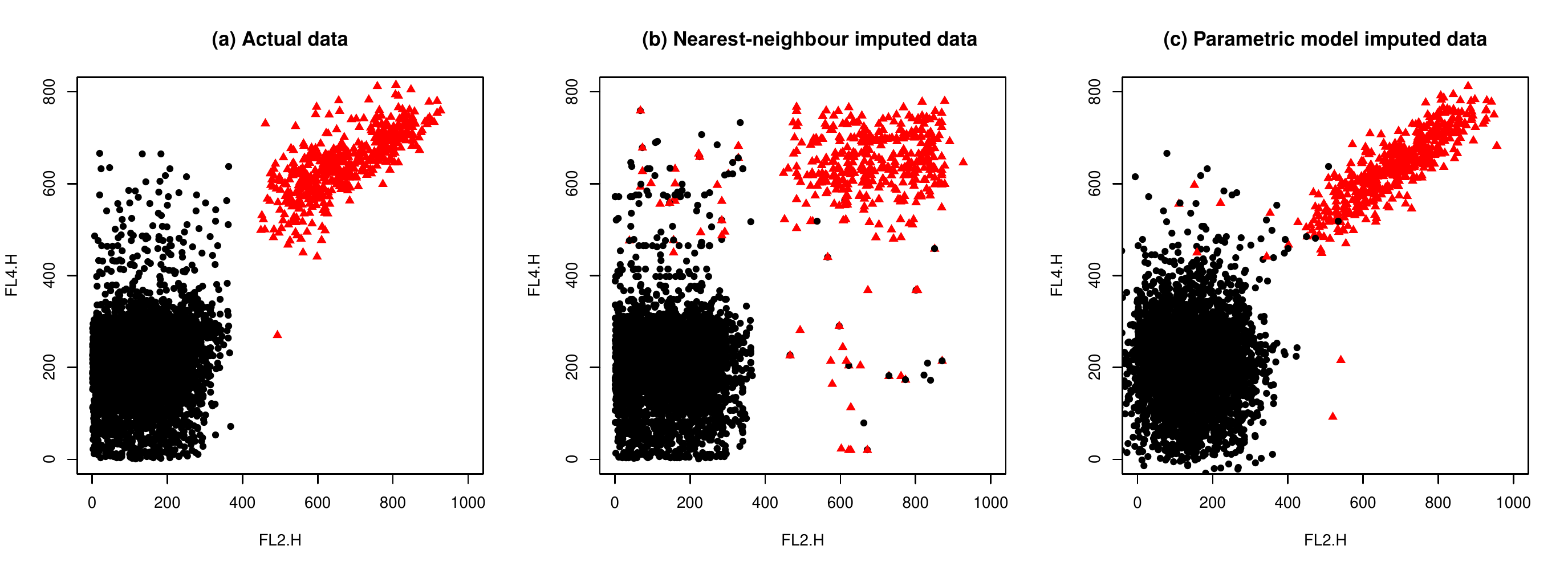}
\caption{Actual and imputed data for the two cell population cytometry dataset in Scenario 2. Black circles and red triangles denote observations from cell subpopulations 1 and 2 respectively. }
\label{fig:cytometry_mixture_scenario2}
\end{figure}

Figure \ref{fig:cytometry_mixture_scenario2} compares the imputed data to the actual and data in Scenario two. Groupwise correlation estimates are also reported in Table \ref{tab:cytometry_correlations}. In Scenario 2, the matching variable FL3.H is able to separate the two cell subpopulations. As such, nearest-neighbour matching produces the correct number of clusters, however the correlation estimate in component two ($\widehat{\rho}_{YZ} =0.13$) is much lower than the true correlation in the source dataset ($\rho_{YZ}=0.74$).  Parametric imputation also produces two clusters, but seems to preserve more of the distributional shape than the nearest-neighbour approach. Nearest-neighbour matching gives very small correlation estimates for each component. Parameteric imputation gives a good estimate of the correlation in subpopulation 2, but underestimates the correlation in subpopulation 1. It is interesting to compare the results in the two scenarios. Both parametric imputation and nearest-neighbour matching produces different correlation estimates in each scenario. The quality of results appears to be sensitive to the choice of matching variable. The matching variable in Scenario 1 is not informative for clustering and this appears to impact the faithfulness of the imputed data for both methods.

\begin{table}[h]
\centering
\begin{tabular}{lrrrr}
\toprule
 &  \multicolumn{2}{c}{Scenario 1} &  \multicolumn{2}{c}{Scenario 2}  \\
 \cmidrule(lr){2-3}   \cmidrule(lr){4-5}
Correlation & Component 1  & Component 2& Component 1 & Component 2 \\ 
\midrule
True sample   &  0.10& 0.74 &  0.10 &  0.74\\
Nearest-neighbour  &-0.04 & -0.69&  0.03 &  0.13 \\
Parameteric Model& 0.16& 0.04   & 0.01 & 0.81 \\
\bottomrule          
\end{tabular}
\caption{Estimates of the $(\vect{Y}, \vect{Z})$ correlation $\rho_{YZ}$ for the two cell subpopulation flow cytometry dataset.}
\label{tab:cytometry_correlations}
\end{table}

\section{Conclusion}
In the statistical matching problem, the identifiability of $f(\vect{x}, \vect{y}, \vect{z}; \vect{\theta})$ does not necessarily imply the conditional independence of $\vect{Y}$ and $\vect{Z}$ given $\vect{X}$. As such, a parametric imputation strategy can give different results to nearest-neighbour imputation. Although nearest-neighbour imputation is a non-parameteric method, it is not assumption free. The nearest-neighbour method makes a conditional independence assumption on the observed random variables that may not be appropriate if the generative model has some latent structure. 

We showed that a model based approach to the statistical matching problem for skew-normal data and Gaussian mixture data can be implemented using data augmentation and the EM algorithm. By reparameterising the complete-data likelihood we can establish a relationship between the statistical matching problem and mixtures of regression models. The reparameterisation is important to give closed form E and M steps. Additionally, we can impose identification constraints on the model to avoid introducing the missing $\vect{Y}$ and $\vect{Z}$ observations into the complete-data log likelihood. This is to minimise the computational expense of each EM iteration. The statistical matching of non-Gaussian data can be significantly more computationally demanding than non-Gaussian data, and this is an important consideration when working with large flow cytometry datasets. 

The artefacts introduced by the nearest-neighbour method can lead to potential issues in downstream analysis, as the imputer's model and the analyst's model should be compatible \citep{rubin_1996_multiple, meng_1994_multiple}. For example, if nearest-neighbour matching introduces spurious clusters, a downstream analyst using mixture models may overestimate the number of subpopulations in the dataset. The conditional independence assumption is not a necessity in order to perform statistical matching, and as such it may be worth exploring alternative non-parametric or semi-parametric imputation methods.

The statistical matching of mixture models is particularly interesting as the group separation on the common $\vect{X}$ variables appears to strongly influence parameter estimation. In some situations, investigators may be able to engineer the data collection process such that informative variables are assigned to the common $\vect{X}$ group. This situation may arise in survey research where respondents are asked different sets of questions, with a smaller set of common questions given to every subject. The selection of common questions $
\vect{X}$ is interesting from an experimental design point of view. 

The statistical matching problem is a challenging data integration scenario that can require specialised algorithms for missing data imputation. We have found that model based approaches can give different results to nonparameteric imputation schemes. However, the pathological nature of the statistical matching problem does limit the accuracy of any imputation scheme. The model based approach can introduce some systemic bias into the imputed data when there is model misspecification. The identification restrictions adopted in this work are all inspired by the conditional independence constraint for the Gaussian distribution $\Sigma_{YZ}=\Sigma_{YZ} = {\Sigma}_{YX}{\Sigma}_{XX}^{-1}{\Sigma}_{XZ}$. This assumption is quite strong, and is not testable. We have seen in the data applications that there is no guarantee that it will hold on real data. As such, an important future research direction is the identification of alternative parameter constraints that are testable.

\bibliographystyle{biom} 
\bibliography{matching}

\section*{Supporting Information}
Additional supporting information may be found online in the Supporting Information section at the end of the article.

\appendix

\section{Appendix}
\subsection{Conditional skew-normal}
The conditional distributions $f(\vect{y}_{i}| \vect{x}_{i}; \vect{\theta})$ and $f(\vect{z}_{i}| \vect{x}_{i}; \vect{\theta})$ belong to a generalised family of skew-normal distributions that also have a latent variable representation \citep{arellano_2006_unification, arellano_2006_unified}.  Let $TN(\mu, \sigma^2, a)$ denote a lower truncated normal distribution where $\mu$ and $\sigma^2$ give the mean and variance of the underling normal distribution and $a$ gives the lower truncation bound. We say $U \sim TN(\mu, \sigma^2, a)$ if $U\overset{d}{=}[W|W >a]$ where $W \sim N(\mu, \sigma^2)$.
The conditional distribution of $\vect{Y}$ and $\vect{Z}$ given $\vect{X}$ can be represented as
\begin{align*}
 \begin{bmatrix}
\vect{Y}\\
\vect{Z}
\end{bmatrix}&= \begin{bmatrix}
\vect{\mu}_{Y|X} \\
\vect{\mu}_{Z|X}
\end{bmatrix} + \begin{bmatrix}
\vect{\delta}_{Y|X} \\
\vect{\delta}_{Z|X}
\end{bmatrix}{U}_{X}  + \begin{bmatrix}
\vect{V}_{Y|X} \\
\vect{V}_{Z|X}
\end{bmatrix},
\end{align*}
where ${U}_{X}$ has a truncated normal distribution  $TN(\tau_{X}, \gamma_{X}, 0)$, and $\vect{V}_{Y|X}$ and $\vect{V}_{Z|X}$ are jointly normally distributed. The location parameters are given by
\begin{align*}
\begin{bmatrix}
\vect{\mu}_{Y|X} \\
\vect{\mu}_{Z|X}
\end{bmatrix} &= \begin{bmatrix}
\mat{\mu}_Y\\
\mat{\mu}_Z
\end{bmatrix}  - \begin{bmatrix}
\Sigma_{YX} \\
\Sigma_{ZX}
\end{bmatrix} \Sigma_{XX}^{-1}(\vect{x}-\vect{\mu}_{X}).
\end{align*}
The latent $U_{X}$ is distributed as a truncated normal $TN(\tau_{X}, \gamma_{X}, 0)$ random variable, where
\begin{align*}
\tau_{X} &= \vect{\delta}_{X}^{\mathsf{T}} (\Sigma_{XX}+\vect{\delta}_{X}\vect{\delta}_{X}^{\mathsf{T}}) ^{-1}(\vect{x}-\vect{\mu}_{X}),\\
\gamma_{X} &= 1-\vect{\delta}_{X}^{\mathsf{T}}(\Sigma_{XX}+\vect{\delta}_{X}\vect{\delta}_{X}^{\mathsf{T}}) ^{-1}\vect{\delta}_{X}.
\end{align*}
The latent $\vect{V}_{Y|X}$ and $\vect{V}_{Z|X}$ have a normal distribution,
\begin{align*}
\begin{bmatrix}
\vect{V}_{Y|X} \\
\vect{V}_{Z|X}
\end{bmatrix} \sim N\left(\begin{bmatrix}
\vect{0}  \\
\vect{0}
\end{bmatrix}, \begin{bmatrix}
\Sigma_{YY|X} & \Sigma_{YZ|X} \\
\Sigma_{ZY|X} & \Sigma_{ZZ|X}
\end{bmatrix} \right).
\end{align*}
The conditional covariance matrix is
\begin{align*}
\begin{bmatrix}
\Sigma_{YY|X} & \Sigma_{YZ|X} \\
\Sigma_{ZY|X} & \Sigma_{ZZ|X}
\end{bmatrix} &= \begin{bmatrix}
{\Sigma}_{YY} &{\Sigma}_{YZ} \\
{\Sigma}_{YZ}  & {\Sigma}_{ZZ} 
\end{bmatrix} -\begin{bmatrix}
{\Sigma}_{YX} \\
{\Sigma}_{ZX}
\end{bmatrix}(\Sigma_{XX}+\vect{\delta}_{X}\vect{\delta}_{X}^{\mathsf{T}}) ^{-1}\begin{bmatrix}
{\Sigma}_{XY} & \Sigma_{XZ}
\end{bmatrix} - \dfrac{1}{\gamma_{X}}\vect{\delta}_{X}\vect{\delta}_{X}^{\mathsf{T}}.
\end{align*}
The conditional skewness parameters are
\begin{align*}
\begin{bmatrix}
\vect{\delta}_{Y|X} \\
\vect{\delta}_{Z|X}
\end{bmatrix} &= \dfrac{1}{\gamma_{X}} \left(\begin{bmatrix}
\vect{\delta}_Y\\
\vect{\delta}_Z
\end{bmatrix}  - \begin{bmatrix}
{\Sigma}_{YX} \\
{\Sigma}_{ZX}
\end{bmatrix} (\Sigma_{XX}+\vect{\delta}_{X}\vect{\delta}_{X}^{\mathsf{T}}) ^{-1} \vect{\delta}_{X}\right).
\end{align*}
\subsection{Asymptotic analysis of skew-normal}
\setcounter{equation}{0}
\renewcommand\theequation{A.\arabic{equation}}
 The hierarchical representation of a skew-normal random variable is useful to analyse the asymptotic behaviour of the nearest-neighbour method for skew-normal data. The asymptotic form of the nearest-neighbour imputation distribution \eqref{eq:nn_asymptotic} will involve the true conditional distributions $f(\vect{y}_{i}| \vect{x}_{i}; \vect{\theta})$ and $f(\vect{z}_{i}| \vect{x}_{i}; \vect{\theta})$. The true marginal $(\vect{Y}_{i}, \vect{Z}_{i})$ distribution can be expressed through a hierarchical model
\begin{align}
\vect{Y}_{i}, \vect{Z}_{i} | \vect{X}_{i}=\vect{x}_{i}, {U}_{X}=u_{X} &\sim N \left( \begin{bmatrix}
\vect{\mu}_{Y|X} \\
\vect{\mu}_{Z|X}
\end{bmatrix}  + \begin{bmatrix}
\vect{\delta}_{Y|X} \\
\vect{\delta}_{Z|X}
\end{bmatrix}{u}_{X}, \begin{bmatrix}
\Sigma_{YY|X} & \Sigma_{YZ|X} \\
\Sigma_{ZY|X} & \Sigma_{ZZ|X}
\end{bmatrix}  \right),   \label{eq:sn_true_yz} \\
U_{X} | \vect{X}_{i}=\vect{x}_{i} &\sim TN(\tau_{X}, \gamma_{X}, 0), \label{eq:sn_true_latent} \\
\vect{X}_{i} &\sim f(\vect{x}_{i}; \vect{\mu}_{X}, \Sigma_{XX}, \vect{\delta}_{X}). \label{eq:sn_true_x}
\end{align}
We can also characterise the asymptotic nearest-neighbour imputation distribution through a hierarchical model. Asymptotically, the effective model used by nearest-neighbour matching is that conditional on $\vect{X}_{i}=\vect{x}_{i}$,
\begin{align}
\vect{Z}_{i} | \vect{X}_{i}=\vect{x}_{i}, {U}_{X}^{(1)}=u_{X}^{(1)} &\sim N \left(
\vect{\mu}_{Z|X} + 
\vect{\delta}_{Z|X}
{u}_{X}^{(1)},  \Sigma_{ZZ|X} \right),  \label{eq:sn_nn_z}  \\
\vect{Y}_{i} | \vect{X}_{i}=\vect{x}_{i}, {U}_{X}^{(2)}=u_{X}
^{(2)}&\sim N \left(
\vect{\mu}_{Y|X} + 
\vect{\delta}_{Y|X}
{u}_{X}^{(2)},  \Sigma_{YY|X} \right),  \label{eq:sn_nn_y} \\
U_{X}^{(1)}, U_{X}^{(2)} | \vect{X}_{i}=\vect{x}_{i} &\overset{i.i.d}{\sim} TN(\tau_{X}, \gamma_{X}, 0), \label{eq:sn_nn_latent} \\
\vect{X}_{i} &\sim f(\vect{x}_{i}; \vect{\mu}_{X}, \Sigma_{XX}, \vect{\delta}_{X}). \label{eq:sn_nn_x}
\end{align}
Comparing  to \eqref{eq:sn_nn_latent} to \eqref{eq:sn_true_latent} we see that an important difference between the nearest-neighbour imputation distribution and the true marginal distribution is the use of two independent latent scaling variables ${U}_{X}^{(1)}$ and ${U}_{X}^{(2)}$ instead of the single latent variable ${U}_{X}$ that appears in the true model. A second difference is that $\vect{Z}_{i}$ and $\vect{Y}_{i}$ are sampled independently conditional on $\vect{X}_{i}$. Levels \eqref{eq:sn_nn_z} and \eqref{eq:sn_nn_y} are independent conditional on the lower layers. The conditional covariance $\Sigma_{YZ|X}$ that appears in the true generative process \eqref{eq:sn_true_latent} does not influence the nearest-neighbour imputation distribution. With these systematic differences, it is reasonable to expect that nearest-neighbour imputation will underestimate the level of dependence between the $\vect{Y}$ and $\vect{Z}$ variables. If we believe that the skew-normal model \eqref{eq:skew_normal_latent_model} is appropriate, it is difficult to justify the use of nearest-neighbour matching as the asymptotic behaviour is not consistent with our initial assumptions.  
\section*{}
\subsection{Asymptotic analysis of Gaussian mixture}
The true joint distribution on the $(\vect{Y}_{i}, \vect{Z}_{i})$ variables has the hierarchical representation:
\begin{align}
\vect{Y}_{i},\vect{Z}_{i} | \vect{X}_{i}=\vect{x}_{i},S_{i}=h &\sim f(\vect{y}_{i}, \vect{z}_{i} | \vect{x}_{i} ; \vect{\theta}_{h}),  \label{eq:true_z_mixture} \\
S_{i} | \vect{X}_{i} = \vect{x}_{i}& \sim \text{Categorical}(\tau_{1}, \ldots, \tau_{g}), \\
\vect{X}_{i} &\sim f(\vect{x}_{i}; \vect{\Psi}),  \label{eq:true_xy_mixture} 
\end{align}
where for $j=1, \ldots, g$:
\begin{align}
\tau_{j} &= \dfrac{\pi_{j}f(\vect{x}_{i} ; \vect{\theta}_{j})}{\sum_{k=1}^{g}\pi_{k}f(\vect{x}_{i}; \vect{\theta}_{k})}. \label{eq:tau_definition}
\end{align}
Although conventional to include the cluster indicator $S_{i}$ at the bottom of the hierarchy, for the purposes of our analysis it is beneficial to sample it conditional on $\vect{X}_{i}=\vect{x}_{i}$. We again analyse the asymptotic behaviour nearest-neighbour matching by determining the form of the asymptotic imputation distribution $g(\vect{x}_{i}, \vect{y}_{i}, \vect{z}_{i})$ \eqref{eq:nn_asymptotic}. The true conditional distributions $f(\vect{y}_{i}| \vect{x}_{i}; \vect{\theta})$ and $f(\vect{z}_{i}| \vect{x}_{i}; \vect{\theta})$ can also be expressed as finite mixture models. In full,
 \begin{align*}
f(\vect{y}_{i} | \vect{x}_{i}; \vect{\Psi}) &= \sum_{j=1}^{g}\tau_{j}f(\vect{y}_{i} | \vect{x}_{i}; \vect{\theta}_{j}) \\
f(\vect{z}_{i} | \vect{x}_{i}; \vect{\Psi}) &= \sum_{j=1}^{g}\tau_{j}f(\vect{z}_{i} | \vect{x}_{i}; \vect{\theta}_{j})
 \end{align*}
 where the weights ${\tau}_{1}, \ldots, \tau_{g}$ are given by \eqref{eq:tau_definition}. Asymptotically speaking, the nearest-neighbour imputation distribution $g(\vect{x}_{i}, \vect{y}_{i}, \vect{z}_{i})$ has the hierarchical representation:
\begin{align}
\vect{Z}_{i} | \vect{X}_{i}=\vect{x}_{i},T_{i}=j &\sim f(\vect{z}_{i} | \vect{x}_{i}; \vect{\theta}_{j}) \label{eq:nn_t_mixture} \\
\vect{Y}_{i} | \vect{X}_{i}=\vect{x}_{i},S_{i}=h &\sim f(\vect{y}_{i} | \vect{x}_{i}; \vect{\theta}_{h}),  \label{eq:nn_s_mixture} \\
S_{i}, T_{i} | \vect{X}_{i}=\vect{x}_{i}& \overset{i.i.d}{\sim} \text{Categorical}(\tau_{1}, \ldots, \tau_{g}) \label{eq:nn_indicator_sampling} \\
\vect{X}_{i} &\sim f(\vect{x}_{i}; \vect{\Psi}),  \label{eq:true_xdy_mixture} 
\end{align}
An important difference between the nearest-neighbour imputation distribution and the true conditional model is the introduction of a second latent variable $T_{i}$ \eqref{eq:nn_indicator_sampling}. If the sampled $T_{i}$ is not equal to the sampled $S_{i}$ at level \eqref{eq:nn_indicator_sampling}, the $\vect{Y}_{i}$ and $\vect{Z}_{i}$ random variables will not be drawn from the same component distribution in levels  \eqref{eq:nn_s_mixture} and \eqref{eq:nn_t_mixture}. The probability of a mismatch is related to how reliably the $\vect{X}$ variables can be used to classify observations, entering through the $\tau_{j}$ terms \eqref{eq:tau_definition}. The imputed joint distribution $g(\vect{y}, \vect{z})$ can introduce spurious clusters if the $\vect{X}$ variables are not useful for clustering the observations.

\section*{Web Appendix:Introduction}
Here we give the E-step and M-step for the EM algorithms for the statistical matching of non-Gaussian data. Recall the hierarchical model for the skew normal distribution:
\begin{align}
\begin{bmatrix}
\vect{X}_{i} \\
\vect{Y}_{i} \\
\vect{Z}_{i}
\end{bmatrix} &= \begin{bmatrix}
\vect{\mu}_{X} \\
\vect{\mu}_{Y} \\
\vect{\mu}_{Z}
\end{bmatrix} + \begin{bmatrix}
\vect{\delta}_{X} \\
\vect{\delta}_{Y} \\
\vect{\delta}_{Z}
\end{bmatrix}{U}_{i}  + \vect{V}_{i} \label{eq:skew_normal_latent_model_app}
\end{align}
where ${U}_{i} \sim TN(0, 1, 0)$ and 
\begin{align*}
\vect{V}_{i} &\sim N\left(
\vect{\mu} = \begin{bmatrix} \vect{0} \\ \vect{0} \\ \vect{0} \end{bmatrix}, {\Sigma} = 
\begin{bmatrix}
{\Sigma}_{XX} & {\Sigma}_{XY} & {\Sigma}_{XZ} \\
{\Sigma}_{YX} & {\Sigma}_{YY} & {\Sigma}_{YZ} \\
{\Sigma}_{ZX} &  {\Sigma}_{ZY} &{\Sigma}_{ZZ}
\end{bmatrix} \right).
\end{align*}
The combined data matrix from dataset $A$ and dataset $B$ is partitioned as
\begin{equation*}
  \left(\begin{array}{cccc}
\shadecell \vect{x}_{1} & \shadecell \vect{y}_{1} & \vect{z}_{1} \\
 \shadecell   \vdots & \shadecell \vdots & \vdots \\
 \shadecell    \vect{x}_{n_{A}} & \shadecell \vect{y}_{n_{A}} & \vect{z}_{n_{A}}\\
  \shadecell   \vect{x}_{n_{A}+1} & \vect{y}_{n_{A}+1} & \shadecell \vect{z}_{n_{A}+1}\\
 \shadecell    \vdots & \vdots & \shadecell \vdots \\
   \shadecell  \vect{x}_{n_{A}+n_{B}} & \vect{y}_{n_{A}+n_{B}} &\shadecell  \vect{z}_{n_{A}+n_{B}}
  \end{array}\right),
\end{equation*}
where shaded cells are observed and blank cells are missing.
\section*{Web Appendix A: Skew-normal distribution}
\subsection*{E-step}
Using the hierarchical model for the skew normal distribution \eqref{eq:skew_normal_latent_model}, define the following conditional expectations given the current parameter estimates:
\begin{align*}
    e_{1,i} &= \begin{cases}
    \mathbb{E}[U_{i} \vert  \vect{x}_{i}, \vect{y}_{i} ] &  \mbox{if } i=1, \ldots, n_{A}, \\
     \mathbb{E}[U_{i} \vert  \vect{x}_{i}, \vect{z}_{i} ] & \mbox{if } i=n_{A}+1, \ldots, n
    \end{cases} \\
    e_{2,i} &= \begin{cases}
    \mathbb{E}[U_{i}^2 \vert  \vect{x}_{i}, \vect{y}_{i} ] &  \mbox{if } i=1, \ldots, n_{A}, \\
     \mathbb{E}[U_{i}^2 \vert  \vect{x}_{i}, \vect{z}_{i} ] & \mbox{if } i=n_{A}+1, \ldots, n.
    \end{cases}
\end{align*}
Let
\begin{align*}
    m_{i} &=\begin{cases} \begin{bmatrix}
    \vect{\delta}_{X}^{\T} &
    \vect{\delta}_{Y}^{\T}
            \end{bmatrix} \left(\begin{bmatrix}
            \Sigma_{XX} & \Sigma_{XY} \\
            \Sigma_{YX} & \Sigma_{YY}
            \end{bmatrix} + \begin{bmatrix} 
            \vect{\delta}_{X} \\
            \vect{\delta}_{Y}
            \end{bmatrix}\begin{bmatrix} 
            \vect{\delta}_{X}^{\T} & 
            \vect{\delta}_{Y}^{\T}
            \end{bmatrix}\right)^{-1}\left(\begin{bmatrix}
            \vect{x}_{i} \\
            \vect{y}_{i}
            \end{bmatrix} - \begin{bmatrix}
            \vect{\mu}_{X} \\
            \vect{\mu}_{Y}
            \end{bmatrix}\right) & \mbox{if } i=1, \ldots, n_{A} \\ \\
            \begin{bmatrix}
    \vect{\delta}_{X}^{\T} &
    \vect{\delta}_{Z}^{\T}
            \end{bmatrix} \left(\begin{bmatrix}
            \Sigma_{XX} & \Sigma_{XZ} \\
            \Sigma_{ZX} & \Sigma_{ZZ}
            \end{bmatrix} + \begin{bmatrix} 
            \vect{\delta}_{X} \\
            \vect{\delta}_{Z}
            \end{bmatrix}\begin{bmatrix} 
            \vect{\delta}_{X}^{\T} & 
            \vect{\delta}_{Z}^{\T}
            \end{bmatrix}\right)^{-1}\left(\begin{bmatrix}
            \vect{x}_{i} \\
            \vect{z}_{i}
            \end{bmatrix} - \begin{bmatrix}
            \vect{\mu}_{X} \\
            \vect{\mu}_{Z}
            \end{bmatrix}\right) & \mbox{if } i=n_{A}+1, \ldots, n_{A}+n_{B}  
\end{cases}     
\end{align*} 
Furthermore define the constants
\begin{align*}
    c_{1} &= \left(1-\begin{bmatrix}
    \vect{\delta}_{X}^{\T} & \vect{\delta}_{Y}^{\T}
    \end{bmatrix}\left(\begin{bmatrix}
            \Sigma_{XX} & \Sigma_{XY} \\
            \Sigma_{YX} & \Sigma_{YY}
            \end{bmatrix} + \begin{bmatrix} 
            \vect{\delta}_{X} \\
            \vect{\delta}_{Y}
            \end{bmatrix}\begin{bmatrix} 
            \vect{\delta}_{X}^{\T} & 
            \vect{\delta}_{Y}^{\T}
            \end{bmatrix}\right)^{-1}\begin{bmatrix}
            \vect{\delta}_{X} \\
            \vect{\delta}_{Y}
            \end{bmatrix} \right) \\
    c_{2} &= \left(1-\begin{bmatrix}
    \vect{\delta}_{X}^{\T} & \vect{\delta}_{Z}^{\T}
    \end{bmatrix}\left(\begin{bmatrix}
            \Sigma_{XX} & \Sigma_{XZ} \\
            \Sigma_{ZX} & \Sigma_{ZZ}
            \end{bmatrix} + \begin{bmatrix} 
            \vect{\delta}_{X} \\
            \vect{\delta}_{Z}
            \end{bmatrix}\begin{bmatrix} 
            \vect{\delta}_{X}^{\T} & 
            \vect{\delta}_{Z}^{\T}
            \end{bmatrix}\right)^{-1}\begin{bmatrix}
            \vect{\delta}_{X} \\
            \vect{\delta}_{Z}
            \end{bmatrix} \right)
\end{align*}
Let $\phi(\cdot)$ denote the standard normal density function and let $\Phi(\cdot)$ denote the  cumulative distribution function of the standard normal distribution. We have the closed form expressions for the expectations \citep{pyne2009automated}.
\begin{align*}
    e_{1,i} &= \begin{cases}
     m_{i} + c_{1}[\phi(m_{i}/c_{1})/\Phi(m_{i}/c_{1})] & \mbox{if } i=1, \ldots, n_{A} \\
     m_{i} + c_{2}[\phi(m_{i}/c_{2})/\Phi(m_{i}/c_{2})]  & \mbox{if } i=n_{A}+1, \ldots, n_{A}
    \end{cases} \\ \\
    e_{2,i} &= \begin{cases} m_{i}^2 + c_{1}^2+c_{1}m_{i}[\phi(m_{i}/c_{1})/\Phi(m_{i}/c_{1})] &  \mbox{if } i=1, \ldots, n_{A} \\
    m_{i}^2 + c_{2}^2+c_{2}m_{i}[\phi(m_{i}/c_{2})/\Phi(m_{i}/c_{2})] &  \mbox{if } i=n_{A}+1, \ldots, n_{A}
    \end{cases}
\end{align*}
\subsection*{M-step}
For $i=1, \ldots, n$ let 
\begin{align*}
    P_{i} &= \left(\vect{x}_{i} - \vect{\mu}_{X}\right)(\vect{x}_{i} - \vect{\mu}_{X})^{\T} - e_{1,i}(\vect{\delta}_{X}(\vect{x}_{i}-\vect{\mu}_{X})^{\T})-((\vect{x}_{i}-\vect{\mu}_{X})\vect{\delta}_{X}^{\T})e_{1,i}+e_{2,i}\vect{\delta}_{X}\vect{\delta}_{X}^{\T} \\
    D_{i} &= e_{1,i}(\vect{x}_{i}-\vect{\mu}_{X}) \\
    L_{i} &= (\vect{x}_{i} - e_{1,i}\vect{\delta}_{X}). 
\end{align*}
The maximisation of the $X$ parameters follows from existing results on the skew normal distribution \citep{pyne2009automated}:
\begin{align*}
    \widehat{\Sigma}_{XX} &= \left(\sum_{i=1}^{n}P_{i}\right) \\
    \widehat{\mu}_{X} &= \left(\sum_{i=1}^{n}L_{i}\right) \\
        \widehat{\delta}_{X} &= \left(\sum_{i=1}^{n}D_{i}\right)\left\lbrace \sum_{i=1}^{n}e_{2,i} \right\rbrace^{-1}.
\end{align*}
Recall that $\Gamma_{A}$ and $\Gamma_{B}$ contain the regression parameters for the conditional regression models: $\Gamma_{A}=[\vect{\alpha}_{Y}\ \vect{\lambda}_{Y} \ \vect{\beta}_{Y}]^{\T}$ and $\Gamma_{B}=[\vect{\alpha}_{Z} \ \vect{\lambda}_{Z} \  \vect{\beta}_{Z}]^{\T}$. The regressions for dataset A and dataset B can be written as
\begin{align}
\mat{Y}_{A} &= \mat{B}_{A}\Gamma_{A} + \vect{\epsilon}_{A},  \label{eq:sn_joint_y_regression_model_app}\\
\mat{Z}_{B} &= \mat{B}_{B}\Gamma_{B} + \vect{\epsilon}_{B}, \label{eq:sn_joint_z_regression_model_app}
\end{align}
where $\vect{\epsilon}_{A} \sim MN(\vect{I}_{n_{A}}, \Omega_{Y})$ and $\vect{\epsilon}_{B} \sim MN(\vect{I}_{n_{B}}, \Omega_{Z})$. The design matrices for the regressions now include the latent $U_{i}$ terms. The complete-data design matrices are given by
\begin{align}
\mat{B}_{A} &= \begin{bmatrix}
1 & u_{1} & \vect{x}_{1}^{\mathsf{T}} \\
1 & u_{2} &\vect{x}_{2}^{\mathsf{T}} \\
\vdots \\
1 &u_{n_{A}} &\vect{x}_{n_{A}}^{\mathsf{T}}
\end{bmatrix}, \quad
\mat{B}_{B} = \begin{bmatrix}
1 &u_{n_{A}+1} &\vect{x}_{n_{A}+1}^{\mathsf{T}} \\
1 &u_{n_{A}+2} &\vect{x}_{n_{A}+2}^{\mathsf{T}} \\
\vdots \\
1 &u_{n_{A}+n_{B}} &\vect{x}_{n_{A}+n_{B}}^{\mathsf{T}}
\end{bmatrix}. \label{eq:design_matrix_sn_app}
\end{align}

We need to calculate the conditional expectations $\mathbb{E}[\mat{B}^{\T}_{A}\mat{B}_{A}]$ and $\mathbb{E}[\mat{B}^{\T}_{B}\mat{B}_{B}]$  given the current parameter estimates and the observed data. For $i=1,\ldots, n$ let
\begin{align*}
    \mat{G}_{i} &= \begin{bmatrix}
    1  \\ e_{1,i} \\ \vect{x}_{i}
    \end{bmatrix}\begin{bmatrix}
    1  & e_{1,i} & \vect{x}_{i}^{\T}
    \end{bmatrix} +  \begin{bmatrix}
    0 & 0 & \vect{0} \\
    0 & e_{2,i}-e_{1,i}^2 & \vect{0} \\
    \vect{0} & \vect{0} & \vect{0} 
    \end{bmatrix}. 
\end{align*}
From standard results on multiple outcome regression models\citep{rencher_methods_2012}, we have a closed form M-step for the regression parameters
\begin{align*}
    \widehat{\Gamma}_{Y} &= \left(\sum_{i=1}^{n_{A}}\mat{G
    }_{i}\right)^{-1}\left(\sum_{i=1}^{n_{A}}  \begin{bmatrix}
    1   \\ e_{1,i}  \\ \vect{x}_{i}
    \end{bmatrix}\vect{y}_{i}^{\T}\right) \\
        \widehat{\Gamma}_{X} &= \left(\sum_{i=n_{A}+1}^{n}\mat{G
    }_{i}\right)^{-1}\left(\sum_{i=n_{A}+1}^{n}  \begin{bmatrix}
    1   \\ e_{1,i}  \\ \vect{x}_{i}
    \end{bmatrix}\vect{z}_{i}^{\T}\right).
\end{align*}
For $i=1, \ldots, n$ let
\begin{align*}
    \mat{R}_{i} &= \begin{cases} \vect{y}_{i}\vect{y}_{i}^{\T}-\vect{y}_{i}\begin{bmatrix}
    1 & e_{1,i} & \vect{x}_{i}^{\T}
    \end{bmatrix}\widehat{\Gamma}_{Y} - \widehat{\Gamma}_{Y}^{\T}\begin{bmatrix}
    1 \\
    e_{1,i} \\
    \vect{x}_{i}
    \end{bmatrix}\vect{y}_{i}^{\T}+\widehat{\Gamma}_{Y}^{\T}G_{i}\widehat{\Gamma}_{Y}, \\ \\
    \vect{z}_{i}\vect{z}_{i}^{\T}-\vect{z}_{i}\begin{bmatrix}
    1 & e_{1,i} & \vect{x}_{i}^{\T}
    \end{bmatrix}\widehat{\Gamma}_{Z} - \widehat{\Gamma}_{Z}^{\T}\begin{bmatrix}
    1 \\
    e_{1,i} \\
    \vect{x}_{i}
    \end{bmatrix}\vect{z}_{i}^{\T}+\widehat{\Gamma}_{Z}^{\T}G_{i}\widehat{\Gamma}_{Z}.
    \end{cases}
\end{align*}
The M-step for the error covariance matrices in the regressions \eqref{eq:sn_joint_y_regression_model_app} and \eqref{eq:sn_joint_z_regression_model_app} is as follows:
\begin{align*}
\widehat{\Omega}_{Y} &= n_{A}^{-1}\left( \sum_{i=1}^{n_{A}}\mat{R}_{i} \right) \\
\widehat{\Omega}_{Z} &= n_{B}^{-1}\left( \sum_{i=n_{A}+1}^{n}\mat{R}_{i} \right).
\end{align*}
The estimates in terms of the original parameters of the skew-normal density can be obtained by substituting into the following equations:
\begin{align*}
    \vect{\mu}_{Y} &= \vect{\alpha}_{Y} + \vect{\beta}_{Y}\vect{\mu}_{X} \\
    \vect{\mu}_{Z} &= \vect{\alpha}_{Z} + \vect{\beta}_{Z}\vect{\mu}_{X} \\
    \vect{\delta}_{Y} &= \vect{\lambda}_{Y} + \vect{\beta}_{Z}\vect{\delta}_{X}\\
    \vect{\delta}_{Z} &= \vect{\lambda}_{Z} + \vect{\beta}_{Z}\vect{\delta}_{X} \\
    \Sigma_{YX} &= \mat{\beta}_{Y}\Sigma_{XX}\\
    \Sigma_{ZX} &= \mat{\beta}_{Z}\Sigma_{XX} \\
    \Sigma_{YY} &= \Omega_{Y}+\Sigma_{YX}\Sigma_{XX}^{-1}\Sigma_{YX}^{\T} \\
    \Sigma_{ZZ}&= \Omega_{Z}+\Sigma_{ZX}\Sigma_{XX}^{-1}\Sigma_{ZX}^{\T} \\
    \Sigma_{YZ} &= \Sigma_{YX}\Sigma_{XX}^{-1}\Sigma_{ZX}^{\T}.
\end{align*}

\section*{Web Appendix B: Mixtures of Gaussians}
\subsection*{E-step}
For $h=1, \ldots, g$, define the posterior class probabilities as
\begin{align*}
    \tau_{i,h} &= \begin{cases}
    N\left(\begin{bmatrix}
    \vect{x}_{i} \\
    \vect{y}_{i}
    \end{bmatrix}; \begin{bmatrix}
    \vect{\mu}_{X}^{(h)}\\
    \vect{\mu}_{Y}^{(h)}
    \end{bmatrix}, \begin{bmatrix}
    \Sigma_{XX}^{(h)} & \Sigma_{XY}^{(h)} \\
    \Sigma_{YX}^{(h)} & \Sigma_{YY}^{(h)}
    \end{bmatrix} \right) \left\lbrace \sum_{r=1}^{g}    N\left(\begin{bmatrix}
    \vect{x}_{i} \\
    \vect{y}_{i}
    \end{bmatrix}; \begin{bmatrix}
    \vect{\mu}_{X}^{(r)}\\
    \vect{\mu}_{Y}^{(r)}
    \end{bmatrix}, \begin{bmatrix}
    \Sigma_{XX}^{(r)} & \Sigma_{XY}^{(r)} \\
    \Sigma_{YX}^{(r)} & \Sigma_{YY}^{(r)}
    \end{bmatrix} \right) \right\rbrace^{-1} & \mbox{if } i=1, \ldots, n_{A}. \\ \\
       N\left(\begin{bmatrix}
    \vect{x}_{i} \\
    \vect{z}_{i}
    \end{bmatrix}; \begin{bmatrix}
    \vect{\mu}_{X}^{(h)} \\
    \vect{\mu}_{Z}^{(h)}
    \end{bmatrix}, \begin{bmatrix}
    \Sigma_{XX}^{(h)} & \Sigma_{XZ}^{(h)} \\
    \Sigma_{ZX}^{(h)} & \Sigma_{ZZ}^{(h)}
    \end{bmatrix} \right) \left\lbrace \sum_{r=1}^{g}    N\left(\begin{bmatrix}
    \vect{x}_{i} \\
    \vect{z}_{i}
    \end{bmatrix}; \begin{bmatrix}
    \vect{\mu}_{X}^{(r)}\\
    \vect{\mu}_{Z}^{(r)}
    \end{bmatrix}, \begin{bmatrix}
    \Sigma_{XX}^{(r)} & \Sigma_{XZ}^{(r)} \\
    \Sigma_{ZX}^{(r)} & \Sigma_{ZZ}^{(r)}
    \end{bmatrix} \right) \right\rbrace^{-1} & \mbox{if } i=n_{A}+1, \ldots, n.
    \end{cases}
\end{align*}
\subsection*{M-step}
For the maximisation of the $X$ parameters we follow standard results for Gaussian mixture models.
\begin{align*}
\widehat{\vect{\mu}}_{X}^{(h)} &= \left(\sum_{i=1}^{n}\tau_{i,h}\vect{x}_{i}\right)\left( \sum_{i=1}^{n}\tau_{i,h}\right)^{-1} \\
    \Sigma_{XX}^{(h)} &= \left(\sum_{i=1}^{n}\tau_{i,h}(\vect{x}_{i}-\widehat{\vect{\mu}}_{X}^{(h)})(\vect{x}_{i}-\widehat{\vect{\mu}}_{X}^{(h)})^{\T} \right)\left(\sum_{i=1}^{n}\tau_{i,h}\right).
\end{align*}
Recall the conditional regression specification for dataset A,
\begin{align}
\vect{Y}_{i} | \vect{X}=\vect{x}_{i}, S_{i}=h & \sim N(\vect{\alpha}_{Y}^{(h)} +\vect{\beta}_{Y}^{(h)}\vect{x}_{i}, \Omega_{Y}^{(h)}), \label{eq:mvn_y_conditional_app}
\end{align}
where $\vect{\beta}_{Y}^{(h)} =  \Sigma_{YX}^{(h)}[\Sigma_{XX}^{(h)}]^{-1}$, $\vect{\alpha}_{Y}^{(h)} = \vect{\mu}_{Y}^{(h)} - \vect{\beta}_{Y}^{(h)} \vect{\mu}_{X}^{(h)}$ and $\Omega_{Y}^{(h)} =  \Sigma_{YY}^{(h)} - \Sigma_{YX}^{(h)}[\Sigma_{XX}^{(h)}]^{-1}\Sigma_{XY}^{(h)}$. A similar model applies for dataset B,
\begin{align}
\vect{Z}_{i} | \vect{X}=\vect{x}_{i}, S_{i}=h & \sim N(\vect{\alpha}_{Z}^{(h)} +\vect{\beta}_{Z}^{(h)}\vect{x}_{i}, \Omega_{Z}^{(h)}), \label{eq:mvn_z_conditional_app}
\end{align}
where $\vect{\beta}_{Z}^{(h)} =  \Sigma_{ZX}^{(h)}[\Sigma_{XX}^{(h)}]^{-1}$, $\vect{\alpha}_{Z}^{(h)} = \vect{\mu}_{Z}^{(h)} - \vect{\beta}_{Z}\vect{\mu}_{X}$ and $\Omega_{Z} =  \Sigma_{ZZ} - \Sigma_{ZX}\Sigma_{XX}^{-1}\Sigma_{XZ}$. The maximisation for the conditional regression parameters then follows from existing results on mixtures of regression models \citep{jones_fitting_1992}. For $i=1, \ldots, n$ define:
\begin{align*}
\mat{G}_{i} &=\begin{bmatrix}
1 \\
\vect{x}_{i}
\end{bmatrix}\begin{bmatrix}
1 & \vect{x}_{i}^{\T}
\end{bmatrix}.
\end{align*}
The $M$-step for the regression coefficients is given by:
\begin{align*}
    \widehat{\Gamma}_{Y}^{(h)} &= \left(\sum_{i=1}^{n_{A}}\tau_{i,h}G_{i}\right)^{-1}\left(\sum_{i=1}^{n_{A}}\tau_{i,h}\begin{bmatrix}
1 \\
\vect{x}_{i}
\end{bmatrix}\vect{y}_{i}^{\T} \right),  \quad h=1, \ldots, g.\\
\widehat{\Gamma}_{Z}^{(h)} &= \left(\sum_{i=n_{A}+1}^{n}\tau_{i,h}G_{i}\right)^{-1}\left(\sum_{i=n_{A}+1}^{n}\tau_{i,h}\begin{bmatrix}
1 \\
\vect{x}_{i}
\end{bmatrix}\vect{z}_{i}^{\T} \right). \quad h=1, \ldots, g.
\end{align*}
For the residual error variances in \eqref{eq:mvn_y_conditional_app} and \eqref{eq:mvn_z_conditional_app} we have
\begin{align*}
    \widehat{\Omega}_{Y}^{(h)} &= \dfrac{1}{\sum_{i=1}^{n_{A}}\tau_{i,h}}\sum_{i=1}^{n_{A}}\tau_{i,h}\left(\vect{y}_{i}-\Gamma_{Y}^{(h)\T}\begin{bmatrix}1  \\ \vect{x}_{i}\end{bmatrix}\right)\left(\vect{y}_{i}-\widehat{\Gamma}_{Y}^{(h)\T}\begin{bmatrix}1  \\ \vect{x}_{i}\end{bmatrix}\right)^{\T},  \quad h=1, \ldots, g. \\
        \widehat{\Omega}_{Z}^{(h)} &= \dfrac{1}{\sum_{i=n_{A}+1}^{n}\tau_{i,h}}\sum_{i=n_{A}+1}^{n}\tau_{i,h}\left(\vect{z}_{i}-\Gamma_{Z}^{(h)\T}\begin{bmatrix}1  \\ \vect{x}_{i}\end{bmatrix}\right)\left(\vect{z}_{i}-\widehat{\Gamma}_{Z}^{(h)\T}\begin{bmatrix}1  \\ \vect{x}_{i}\end{bmatrix}\right)^{\T}. \quad h=1, \ldots, g.
\end{align*}
For $h-1, \ldots, g$, the maximum likelihood estimates of the original parameters can then be obtained by substituting into the following formulae:
\begin{align*}
    \vect{\mu}_{Y}^{(h)} &= \vect{\alpha}_{Y}^{(h)} + \vect{\beta}_{Y}^{(h)}\vect{\mu}_{X}^{(h)} \\
    \vect{\mu}_{Z}^{(h)} &= \vect{\alpha}_{Z}^{(h)} + \vect{\beta}_{Z}^{(h)}\vect{\mu}_{X}^{(h)} \\
    \Sigma_{YX}^{(h)} &= \mat{\beta}_{Y}^{(h)}\Sigma_{XX}^{(h)}\\
    \Sigma_{ZX}^{(h)} &= \mat{\beta}_{Z}^{(h)}\Sigma_{XX}^{(h)} \\
    \Sigma_{YY}^{(h)} &= \Omega_{Y}^{(h)}+\Sigma_{YX}^{(h)}[\Sigma_{XX}^{(h)}]^{-1}\Sigma_{YX}^{(h)\T} \\
    \Sigma_{ZZ}^{(h)}&= \Omega_{Z}^{(h)}+\Sigma_{ZX}^{(h)}[\Sigma_{XX}^{(h)}]^{-1}\Sigma_{ZX}^{(h)\T} \\
    \Sigma_{YZ}^{(h)} &= \Sigma_{YX}^{(h)}[\Sigma_{XX}^{(h)}]^{-1}\Sigma_{ZX}^{(h)\T}.
\end{align*}
\section*{Web Appendix C: Mixtures of Skew-normal distributions}
\subsection*{E-step}
For $h=1, \ldots, g$, define the posterior class probabilities as
\newline
\resizebox{\linewidth}{!}{
  \begin{minipage}{\linewidth}
\begin{align*}
    \tau_{i,h} &= \begin{cases}
    f\left(\begin{bmatrix}
    \vect{x}_{i} \\
    \vect{y}_{i}
    \end{bmatrix}; \begin{bmatrix}
    \vect{\mu}_{X}^{(h)}\\
    \vect{\mu}_{Y}^{(h)}
    \end{bmatrix}, \begin{bmatrix}
    \Sigma_{XX}^{(h)} & \Sigma_{XY}^{(h)} \\
    \Sigma_{YX}^{(h)} & \Sigma_{YY}^{(h)}
    \end{bmatrix}, \begin{bmatrix}
    \vect{\delta}_{X}^{(h)} \\
    \vect{\delta}_{Y}^{(h)}
    \end{bmatrix} \right) \left\lbrace \sum_{r=1}^{g}    f\left(\begin{bmatrix}
    \vect{x}_{i} \\
    \vect{y}_{i}
    \end{bmatrix}; \begin{bmatrix}
    \vect{\mu}_{X}^{(r)}\\
    \vect{\mu}_{Y}^{(r)}
    \end{bmatrix}, \begin{bmatrix}
    \Sigma_{XX}^{(r)} & \Sigma_{XY}^{(r)} \\
    \Sigma_{YX}^{(r)} & \Sigma_{YY}^{(r)}
    \end{bmatrix},\begin{bmatrix}
    \vect{\delta}_{X}^{(r)} \\
    \vect{\delta}_{Y}^{(r)}
    \end{bmatrix} \right) \right\rbrace^{-1} & \mbox{if } i=1, \ldots, n_{A} \\ \\
       f\left(\begin{bmatrix}
    \vect{x}_{i} \\
    \vect{z}_{i}
    \end{bmatrix}; \begin{bmatrix}
    \vect{\mu}_{X}^{(h)} \\
    \vect{\mu}_{Z}^{(h)}
    \end{bmatrix}, \begin{bmatrix}
    \Sigma_{XX}^{(h)} & \Sigma_{XZ}^{(h)} \\
    \Sigma_{ZX}^{(h)} & \Sigma_{ZZ}^{(h)}
    \end{bmatrix},\begin{bmatrix}
    \vect{\delta}_{X}^{(h)} \\
    \vect{\delta}_{Z}^{(h)}
    \end{bmatrix} \right) \left\lbrace \sum_{r=1}^{g}    f\left(\begin{bmatrix}
    \vect{x}_{i} \\
    \vect{z}_{i}
    \end{bmatrix}; \begin{bmatrix}
    \vect{\mu}_{X}^{(r)}\\
    \vect{\mu}_{Z}^{(r)}
    \end{bmatrix}, \begin{bmatrix}
    \Sigma_{XX}^{(r)} & \Sigma_{XZ}^{(r)} \\
    \Sigma_{ZX}^{(r)} & \Sigma_{ZZ}^{(r)}
    \end{bmatrix}, \begin{bmatrix}
    \vect{\delta}_{X}^{(r)} \\
    \vect{\delta}_{Z}^{(r)}
    \end{bmatrix}\right) \right\rbrace^{-1} & \mbox{if } i=n_{A}+1, \ldots, n
    \end{cases}
\end{align*}
\end{minipage}
}
\newline
\newline
Where $f(\vect{v}; \vect{\mu}, \Sigma, \vect{\delta})$ denotes the skew normal density with parameters  $\vect{\mu}$, $\Sigma$ and  $\vect{\delta}$ evaluated at the point $\vect{v}$. Using the hierarchical model for the skew normal distribution \eqref{eq:skew_normal_latent_model_app}, define the following conditional expectations given 
\begin{align*}
    e_{1,i,h} &= \begin{cases}
    \mathbb{E}[U_{i} \vert  \vect{x}_{i}, \vect{y}_{i}, S_{i}=h ] &  \mbox{if } i=1, \ldots, n_{A}, \\
     \mathbb{E}[U_{i} \vert  \vect{x}_{i}, \vect{z}_{i}, S_{i}=h] & \mbox{if } i=n_{A}+1, \ldots, n
    \end{cases} \\
    e_{2,i,h} &= \begin{cases}
    \mathbb{E}[U_{i}^2 \vert  \vect{x}_{i}, \vect{y}_{i}, S_{i}=h] &  \mbox{if } i=1, \ldots, n_{A}, \\
     \mathbb{E}[U_{i}^2 \vert  \vect{x}_{i}, \vect{z}_{i}, S_{i}=h] & \mbox{if } i=n_{A}+1, \ldots, n
    \end{cases}
\end{align*}
For $h=1, \ldots, g$, define
\begin{align*}
    m_{i,h} &=\begin{cases} \begin{bmatrix}
    \vect{\delta}_{X}^{(h)\T} &
    \vect{\delta}_{Y}^{(h)\T}
            \end{bmatrix} \left(\begin{bmatrix}
            \Sigma_{XX}^{(h)} & \Sigma_{XY}^{(h)} \\
            \Sigma_{YX}^{(h)} & \Sigma_{YY}^{(h)}
            \end{bmatrix} + \begin{bmatrix} 
            \vect{\delta}_{X}^{(h)} \\
            \vect{\delta}_{Y}^{(h)}
            \end{bmatrix}\begin{bmatrix} 
            \vect{\delta}_{X}^{(h)\T} & 
            \vect{\delta}_{Y}^{(h)\T}
            \end{bmatrix}\right)^{-1}\left(\begin{bmatrix}
            \vect{x}_{i} \\
            \vect{y}_{i}
            \end{bmatrix} - \begin{bmatrix}
            \vect{\mu}_{X}^{(h)} \\
            \vect{\mu}_{Y}^{(h)}
            \end{bmatrix}\right) & \mbox{if } i=1, \ldots, n_{A} \\ \\
            \begin{bmatrix}
    \vect{\delta}_{X}^{(h)\T} &
    \vect{\delta}_{Z}^{(h)\T}
            \end{bmatrix} \left(\begin{bmatrix}
            \Sigma_{XX}^{(h)} & \Sigma_{XZ}^{(h)} \\
            \Sigma_{ZX}^{(h)} & \Sigma_{ZZ}^{(h)}
            \end{bmatrix} + \begin{bmatrix} 
            \vect{\delta}_{X}^{(h)} \\
            \vect{\delta}_{Z}^{(h)}
            \end{bmatrix}\begin{bmatrix} 
            \vect{\delta}_{X}^{(h)\T} & 
            \vect{\delta}_{Z}^{(h)\T}
            \end{bmatrix}\right)^{-1}\left(\begin{bmatrix}
            \vect{x}_{i} \\
            \vect{z}_{i}
            \end{bmatrix} - \begin{bmatrix}
            \vect{\mu}_{X}^{(h)} \\
            \vect{\mu}_{Z}^{(h)}
            \end{bmatrix}\right) & \mbox{if } i=n_{A}+1, \ldots, n_{A}+n_{B}  
\end{cases}     
\end{align*} 
Furthermore define the constants
\begin{align*}
    c_{1,h} &= \left(1-\begin{bmatrix}
    \vect{\delta}_{X}^{(h)\T} & \vect{\delta}_{Y}^{(h)\T}
    \end{bmatrix}\left(\begin{bmatrix}
            \Sigma_{XX}^{(h)} & \Sigma_{XY}^{(h)} \\
            \Sigma_{YX}^{(h)} & \Sigma_{YY}^{(h)}
            \end{bmatrix} + \begin{bmatrix} 
            \vect{\delta}_{X}^{(h)} \\
            \vect{\delta}_{Y}^{(h)}
            \end{bmatrix}\begin{bmatrix} 
            \vect{\delta}_{X}^{(h)\T} & 
            \vect{\delta}_{Y}^{(h)\T}
            \end{bmatrix}\right)^{-1}\begin{bmatrix}
            \vect{\delta}_{X}^{(h)} \\
            \vect{\delta}_{Y}^{(h)}
            \end{bmatrix} \right) \\
    c_{2,h} &= \left(1-\begin{bmatrix}
    \vect{\delta}_{X}^{(h)\T} & \vect{\delta}_{Z}^{(h)\T}
    \end{bmatrix}\left(\begin{bmatrix}
            \Sigma_{XX}^{(h)} & \Sigma_{XZ}^{(h)} \\
            \Sigma_{ZX}^{(h)} & \Sigma_{ZZ}^{(h)}
            \end{bmatrix} + \begin{bmatrix} 
            \vect{\delta}_{X}^{(h)} \\
            \vect{\delta}_{Z}^{(h)}
            \end{bmatrix}\begin{bmatrix} 
            \vect{\delta}_{X}^{(h)\T} & 
            \vect{\delta}_{Z}^{(h)\T}
            \end{bmatrix}\right)^{-1}\begin{bmatrix}
            \vect{\delta}_{X}^{(h)} \\
            \vect{\delta}_{Z}^{(h)}
            \end{bmatrix} \right)
\end{align*}
Let $\phi(\cdot)$ denote the standard normal density function and let $\Phi(\cdot)$ denote the  cumulative distribution function of the standard normal distribution. We have the closed form expressions for the expectations.
\begin{align*}
    e_{1,i,h} &= \begin{cases}
     m_{i,h} + c_{1,h}[\phi(m_{i,h}/c_{1,h})/\Phi(m_{i,h}/c_{1,h})] & \mbox{if } i=1, \ldots, n_{A} \\
     m_{i,h} + c_{2,h}[\phi(m_{i,h}/c_{2,h})/\Phi(m_{i,h}/c_{2,h})]  & \mbox{if } i=n_{A}+1, \ldots, n_{A}
    \end{cases} \\ \\
    e_{2,i,h} &= \begin{cases} m_{i,h}^2 + c_{1,h}^2+c_{1,h}m_{i,h}[\phi(m_{i,h}/c_{1,h})/\Phi(m_{i,h}/c_{1,h})]  & \mbox{if } i=1, \ldots, n_{A} \\
    m_{i,h}^2 + c_{2,h}^2+c_{2,h}m_{i,h}[\phi(m_{i,h}/c_{2,h})/\Phi(m_{i,h}/c_{2,h})]  & \mbox{if } i=n_{A}+1, \ldots, n_{A}
    \end{cases}
\end{align*}
\subsection*{M-step}
For $i=1, \ldots, n$ let 
\begin{align*}
    P_{i,h} &= (\vect{x}_{i} - \vect{\mu}_{X}^{(h)})(\vect{x}_{i} - \vect{\mu}_{X}^{(h)})^{\T} - e_{1,i,h}(\vect{\delta}_{X}^{(h)}(\vect{x}_{i}-\vect{\mu}_{X}^{(h)})^{\T})-((\vect{x}_{i}-\vect{\mu}_{X}^{(h)})\vect{\delta}_{X}^{(h)\T})e_{1,i,h}+e_{2,i,h}\vect{\delta}_{X}^{(h)}\vect{\delta}_{X}^{(h)\T} \\
    D_{i,h} &= e_{1,i,h}(\vect{x}_{i}-\vect{\mu}_{X}^{(h)}) \\
    L_{i,h} &= (\vect{x}_{i} - e_{1,i,h}\vect{\delta}_{X}^{(h)}) 
\end{align*}
The maximisation of the $X$ parameters follows from existing results on mixtures of skew normal distributions \citep{pyne2009automated}. For $h=1, \ldots, g$:
\begin{align*}
    \widehat{\Sigma}_{XX}^{(h)} &= \left(\sum_{i=1}^{n}P_{i,h}\right)\left(\sum_{i=1}^{n}\tau_{i,h} \right)^{-1} \\
    \widehat{\mu}_{X} &= \left(\sum_{i=1}^{n}L_{i,h}\right)\left(\sum_{i=1}^{n}\tau_{i,h} \right)^{-1} \\
        \widehat{\delta}_{X} &= \left(\sum_{i=1}^{n}D_{i,h}\right)\left\lbrace \sum_{i=1}^{n}\tau_{i,h}e_{2,i,h} \right\rbrace^{-1}  \\
\end{align*}
The complete-data model can be expressed in terms of conditional regression specifications. Let $\vect{\beta}_{Y}^{(h)} =  \Sigma_{YX}^{(h)}[\Sigma_{XX}^{(h)}]^{-1},
\vect{\alpha}_{Y}^{(h)} = \vect{\mu}_{Y}^{(h)} - \vect{\beta}_{Y}^{(h)} \vect{\mu}_{X}^{(h)},
\Omega_{Y}^{(h)} =  \Sigma_{YY}^{(h)} - \Sigma_{YX}^{(h)}[\Sigma_{XX}^{(h)}]^{-1}\Sigma_{XY}^{(h)}$ and  $\vect{\lambda}_{Y}^{(h)} = \vect{\delta}_{Y}^{(h)} - \vect{\beta}_{Y}^{(h)}\vect{\delta}_{X}^{(h)}$.
\begin{align}
\vect{Y}_{i} | \vect{X}_{i}=\vect{x}_{i}, U_{i}=u_{i}, S_{i}=h & \sim N(\vect{\alpha}_{Y}^{(h)} + \vect{\lambda}_{Y}^{(h)}u_{i} +\vect{\beta}_{Y}^{(h)}\vect{x}_{i}, \Omega_{Y}^{(h)}) \label{eq:mix_skew_normal_y_regression_app}
\end{align}
Let  $\vect{\beta}_{Z}^{(h)} =  \Sigma_{ZX}^{(h)}[\Sigma_{XX}^{(h)}]^{-1},
\vect{\alpha}_{Z}^{(h)} = \vect{\mu}_{Z}^{(h)} - \vect{\beta}_{Z}^{(h)} \vect{\mu}_{X}^{(h)}, 
\Omega_{Z}^{(h)} =  \Sigma_{ZZ}^{(h)} - \Sigma_{ZX}^{(h)}[\Sigma_{XX}^{(h)}]^{-1}\Sigma_{XZ}^{(h)}$ and $\vect{\lambda}_{Z}^{(h)} = \vect{\delta}_{Z}^{(h)} - \vect{\beta}_{Z}^{(h)}\vect{\delta}_{X}^{(h)}$. Similarly the conditional distribution of $\vect{Z}_{i}$ given $\vect{X}_{i}$ and the latent scaling variable $U_{i}$ can be represented as a regression model
\begin{align}
\vect{Z}_{i} | \vect{X}_{i}=\vect{x}_{i}, U_{i}=u_{i}, S_{i}=h & \sim N(\vect{\alpha}_{Z}^{(h)} +\vect{\lambda}_{Z}^{(h)}u_{i} +\vect{\beta}_{Z}^{(h)}\vect{x}_{i}, \Omega_{Z}^{(h)}).  \label{eq:mix_skew_normal_z_regresssion_app}
\end{align}
We need to calculate the conditional expectations  of the predictors in the regressions given the current parameter estimates and the observed data. For $i=1, \ldots, nn$ and $h=1, \ldots, g$ let
\begin{align*}
    \mat{G}_{i,h} &= \begin{bmatrix}
    1  \\ e_{1,i,h} \\ \vect{x}_{i}
    \end{bmatrix}\begin{bmatrix}
    1  & e_{1,i,h} & \vect{x}_{i}^{\T}
    \end{bmatrix} +  \begin{bmatrix}
    0 & 0 & \vect{0} \\
    0 & e_{2,i,h}-e_{1,i,h}^2 & \vect{0} \\
    \vect{0} & \vect{0} & \vect{0} 
    \end{bmatrix}. 
\end{align*}
Recall that $\Gamma_{A}^{(h)}$ and $\Gamma_{B}^{(h)}$ contain the regression parameters for component $h$, so $\Gamma_{A}^{(h)}=[\vect{\alpha}_{Y}^{(h)}\ \vect{\lambda}_{Y}^{(h)} \ \vect{\beta}_{Y}^{(h)}]^{\T}$ and $\Gamma_{B}^{(h)}=[\vect{\alpha}_{Z}^{(h)} \ \vect{\lambda}_{Z}^{(h)} \  \vect{\beta}_{Z}]^{\T}$ for $h=1, \ldots, g$. We have a closed form M-step for the $\vect{Y}$ regression parameters in \eqref{eq:mix_skew_normal_y_regression_app} and the $\vect{Z}$ regression parameters parameters in \eqref{eq:mix_skew_normal_z_regresssion_app}:
\begin{align*}
    \widehat{\Gamma}_{Y} &= \left(\sum_{i=1}^{n_{A}}\tau_{i,h}\mat{G
    }_{i,h}\right)^{-1}\left(\sum_{i=1}^{n_{A}} \tau_{i,h} \begin{bmatrix}
    1   \\ e_{1,i,h}  \\ \vect{x}_{i,h}
    \end{bmatrix}\vect{y}_{i}^{\T}\right) \\
        \widehat{\Gamma}_{X} &= \left(\sum_{i=n_{A}+1}^{n}\tau_{i,h}\mat{G
    }_{i,h}\right)^{-1}\left(\sum_{i=n_{A}+1}^{n}\tau_{i,h}  \begin{bmatrix}
    1   \\ e_{1,i,h}  \\ \vect{x}_{i}
    \end{bmatrix}\vect{z}_{i}^{\T}\right)
\end{align*}
For $h=1, \ldots, g$ define:
\begin{align*}
    \mat{R}_{i,h} &= \begin{cases} \vect{y}_{i}\vect{y}_{i}^{\T}-\vect{y}_{i}\begin{bmatrix}
    1 & e_{1,i,h} & \vect{x}_{i,h}^{\T}
    \end{bmatrix}\widehat{\Gamma}_{Y} - \widehat{\Gamma}_{Y}^{\T}\begin{bmatrix}
    1 \\
    e_{1,i,h} \\
    \vect{x}_{i}
    \end{bmatrix}\vect{y}_{i}^{\T}+\widehat{\Gamma}_{Y}^{\T}\mat{G}_{i,h}\widehat{\Gamma}_{Y} & \mbox{if } i=1, \ldots, n_{A} \\ \\
    \vect{z}_{i}\vect{z}_{i}^{\T}-\vect{z}_{i}\begin{bmatrix}
    1 & e_{1,i,h} & \vect{x}_{i}^{\T}
    \end{bmatrix}\widehat{\Gamma}_{Z} - \widehat{\Gamma}_{Z}^{\T}\begin{bmatrix}
    1 \\
    e_{1,i,h} \\
    \vect{x}_{i}
    \end{bmatrix}\vect{z}_{i}^{\T}+\widehat{\Gamma}_{Z}^{\T}\mat{G}_{i,h}\widehat{\Gamma}_{Z} & \mbox{if } i=n_{A}+1, \ldots, n_{A}+n_{B}.
    \end{cases}
\end{align*}
The M-step for the error covariance matrices in the regressions \eqref{eq:mix_skew_normal_y_regression} and \eqref{eq:mix_skew_normal_z_regresssion} is as follows:
\begin{align*}
\widehat{\Omega}_{Y}^{(h)} &= \dfrac{1}{\sum_{i=1}^{n_{A}}\tau_{i,h}} \left( \sum_{i=1}^{n_{A}}\tau_{i,h}\mat{R}_{i,h} \right)  \quad h=1, \ldots, g. \\
\widehat{\Omega}_{Z}^{(h)} &= \dfrac{1}{\sum_{i=n_{A}+1}^{n}\tau_{i,h}}\left( \sum_{i=n_{A}+1}^{n}\tau_{i,h}\mat{R}_{i,h} \right) \quad h=1, \ldots, g.
\end{align*}
The estimates in the original parameterisation for the skew-normal model can be obtained using the following results. For $h=1, \ldots, g$ we have the relationships:
\begin{align*}
    \vect{\mu}_{Y}^{(h)} &= \vect{\alpha}_{Y}^{(h)} + \vect{\beta}_{Y}^{(h)}\vect{\mu}_{X}^{(h)} \\
    \vect{\mu}_{Z}^{(h)} &= \vect{\alpha}_{Z}^{(h)} + \vect{\beta}_{Z}^{(h)}\vect{\mu}_{X}^{(h)} \\
    \vect{\delta}_{Y}^{(h)} &= \vect{\lambda}_{Y}^{(h)} + \vect{\beta}_{Z}^{(h)}\vect{\delta}_{X}^{(h)}\\
    \vect{\delta}_{Z}^{(h)} &= \vect{\lambda}_{Z}^{(h)} + \vect{\beta}_{Z}^{(h)}\vect{\delta}_{X}^{(h)} \\
    \Sigma_{YX}^{(h)} &= \mat{\beta}_{Y}^{(h)}\Sigma_{XX}^{(h)}\\
    \Sigma_{ZX}^{(h)} &= \mat{\beta}_{Z}^{(h)}\Sigma_{XX}^{(h)} \\
    \Sigma_{YY}^{(h)} &= \Omega_{Y}^{(h)}+\Sigma_{YX}^{(h)}[\Sigma_{XX}^{(h)}]^{-1}\Sigma_{YX}^{(h)\T} \\
    \Sigma_{ZZ}^{(h)}&= \Omega_{Z}^{(h)}+\Sigma_{ZX}^{(h)}[\Sigma_{XX}^{(h)}]^{-1}\Sigma_{ZX}^{(h)\T} \\
    \Sigma_{YZ}^{(h)} &= \Sigma_{YX}^{(h)}[\Sigma_{XX}^{(h)}]^{-1}\Sigma_{ZX}^{(h)\T}.
\end{align*}

\end{document}